\documentclass[ aip, jmp, reprint,]{revtex4}
\usepackage{amsfonts}
\usepackage{url}
\usepackage{amssymb}
\usepackage{amsmath}
\usepackage{graphicx}
\usepackage{dcolumn}
\usepackage{bm}

\begin{document}

\title[Kinetic theory of accretion disc plasmas]{\textbf{The kinetic theory of
quasi-stationary collisionless accretion disc plasmas}}
\author{C. Cremaschini}
\affiliation{International School for Advanced Studies (SISSA), Trieste, Italy}
\affiliation{Consortium for Magnetofluid Dynamics, University of Trieste, Trieste, Italy}
\author{J. C. Miller}
\affiliation{Department of Physics (Astrophysics), University of Oxford, Oxford, UK}
\affiliation{International School for Advanced Studies (SISSA), Trieste, Italy}
\author{M. Tessarotto}
\affiliation{Department of Mathematics and Informatics, University of Trieste, Trieste,
Italy}
\affiliation{Consortium for Magnetofluid Dynamics, University of Trieste, Trieste, Italy}
\date{\today }

\begin{abstract}
Astrophysical plasmas in accretion discs are usually treated in
the framework of fluid or MHD approaches but there are some
situations where these treatments become inadequate and one needs
to revert to the more fundamental underlying kinetic theory. This
occurs when the plasma becomes effectively collisionless or
weakly-collisional such as, for example, in radiatively
inefficient accretion flows onto black holes. In this paper, we
lay down the basics of kinetic theory in these contexts. In
particular, we formulate the kinetic theory for quasi-stationary
collisionless accretion disc plasmas in the framework of a
Vlasov-Maxwell description, taking the plasma to be
non-relativistic, axisymmetric, gravitationally-bound and subject
to electromagnetic fields. Quasi-stationary solutions for the
kinetic distribution functions are constructed which are shown to
admit temperature anisotropies. The physical implications of the
theory are then investigated and the equations of state and
angular momentum conservation law are discussed. Analysis of the
Ampere equation reveals the existence of a quasi-stationary
kinetic dynamo which gives rise to self-generation of poloidal and
azimuthal magnetic fields and operates even in the absence of
turbulence and/or instability phenomena.
\end{abstract}

 \maketitle



\section{Introduction}

Fluids are usually treated as continuous media with their time
evolution being determined by a suitable set of fluid equations,
even if the underlying fundamental description is in terms of a
discrete system of particles, whose dynamics is deterministic but
characterized by random initial conditions. A continuum
description for such systems applies when the kinetic description
has a continuous phase-space probability density, given by a
single-particle kinetic distribution function (KDF), which
satisfies a suitable kinetic equation. Once the KDF is prescribed,
all of the continuum fluid moments can be represented in terms of
well-defined constitutive equations, determined via appropriate
velocity moments of the KDF. When binary Coulomb collisions are
negligible (collisionless or weakly-collisional plasmas), one
should return to the kinetic-theory description. In these cases,
\textquotedblleft stand-alone\textquotedblright\ fluid or
magneto-hydrodynamics (MHD) approaches formulated independently of
an underlying kinetic theory can usually provide, at best, a
partial description of the plasma phenomenology. This is because
of two possible inconsistencies which may arise. Firstly, the set
of fluid equations is generally not closed, and so requires the
independent prescription of equations of state which may not give
rise to a self-consistent system. Secondly, in these approaches no
account is usually given of microscopic phase-space particle
dynamics (including single-particle conservation laws) or of
phase-space plasma collective phenomena (kinetic effects). These
issues are naturally addressed within a kinetic treatment, where
the fluid fields are determined from the underlying
self-consistent KDF. This means that both the equations of state
and the constitutive equations for the fluid fields (see Sections
7-9) then follow uniquely from the microscopic dynamics.

This paper is concerned with the description of stationary or
slowly-time-varying (quasi-stationary) phenomena occurring in
astrophysical plasmas, focusing particularly on accretion discs
around compact objects. For matter in the inner parts of such
discs, the plasma may be either collisional (as in the majority of
cases) or collisionless, depending on the circumstances. A notable
example of the latter case is provided by radiatively inefficient
flows (RIAFs) arising in low-density geometrically-thick discs
around black holes (\cite{Narayan}). Matter in these systems is
thought to consist of a two-temperature plasma, with the ion
temperature being much higher than the electron one, and the
Coulomb collision timescale being much longer than the inflow
time. Other interesting applications for collisionless or
weakly-collisional plasmas concern accretion discs around neutron
stars and white dwarfs. In the inner regions of such discs, the
magnetic field of the central object may become dominant and ions
and electrons can be collisionally decoupled so as to sustain
different temperatures. This happens, in particular, if the
radiative cooling time-scale of the electrons is much shorter than
the time-scale for electron-ion collisions so that the electrons
and ions are thermally decoupled and can have different
temperatures (\cite{Saxton2005,Saxton2007}).

Treatments of accretion discs have often been made in terms of a
purely fluid-dynamical approach to which was added an
\textquotedblleft anomalous\textquotedblright\ form of viscosity
(i.e. one not due to binary particle collisions), following simple
intuitive turbulence models such as that giving rise to $\alpha
$-discs. However, it is now almost universally believed that the
origin of the effective viscosity lies in magnetic phenomena (such
as the magneto-rotational instability, MRI) and that the medium
needs to be treated as a magnetized plasma, when making detailed
investigations, rather than as a simple unmagnetized neutral
fluid. Almost always, these calculations are then performed within
the stand-alone continuum MHD treatments. However, as indicated
above, the underlying plasma kinetic theory needs to be directly
considered when treating collisionless or weakly-collisional
plasmas. An approach of this type has been presented in two recent
papers (\cite{Cr2010a,Cr2011}, hereafter referred to as Paper I
and Paper II respectively).

The aim of the present paper is to address in detail the
astrophysical aspects of the theory previously developed. We focus
on accretion discs composed of collisionless plasma which can be
subject to strong electro-magnetic (EM) fields and which are
gravitationally bound, in the sense of being confined by the
combined gravitational potential of the central compact object and
the disc. We deal with plasma which is sufficiently far from the
central object so that it can be treated non-relativistically,
with both $k_{B}T/m_{s}c^{2}$ and $(R\Omega /c)^{2}$ being $<0.1$,
where $k_{B}$ is the Boltzmann constant, $T$ is the temperature,
$m_{s}$ is the rest-mass of the species of plasma particle
labelled with the index $s$ ($s=e,i$ for an electron-ion plasma),
$R$ is the cylindrical radial distance from the central object
($z$ will be the vertical cylindrical coordinate), and $\Omega $
is the local angular velocity in the disc. We do not include any
interactions with radiation since we are not considering any
background radiation field and radiation-reaction effects
(associated with the self-emission of radiation) are negligible
for our non-relativistic plasma. We envisage magnetic fields,
on the other hand, which can in principle be either \emph{external} ($%
\mathbf{B}^{ext}$), or \emph{self-generated} within the disc itself ($%
\mathbf{B}^{self}$). Estimated values coming from observations of
accretion
discs around compact objects include magnetic-field strengths in the range $%
B\sim 10^{1}-10^{8}G$ (\cite{Frank,Vietri}). For
these systems, estimates for species temperatures usually lie in the ranges $%
T_{i}\sim 10^{4}-10^{12}K$ and $T_{e}\sim 10^{4}-10^{8}K$\textbf{\
}for ions and electrons respectively. Particle densities in
different types of accretion disc around compact objects span a
very wide range of values, but here we focus on the bottom end of
it. In particular, for discussing the collisionless regime of
non-relativistic AD plasmas, the number density is taken to be in
the interval $n_{s}\sim 10^{6}-10^{15}cm^{-3}$. In the case of
hydrogen-ion systems, the Spitzer collision time $\tau _{Ci}$, the
Larmor rotation time $\tau _{Li}$ and the Langmuir time $\tau
_{pi}$ are then respectively in the ranges $\tau _{Ci}\simeq
10^{-2}-10^{19}s,$ $\tau _{Li}\simeq 10^{-4}-10^{-11}s$ and $\tau
_{pi}\simeq 1-10^{-4}s$, while the corresponding estimates for the
Debye length $\lambda _{D}$, the mean free
path $\lambda _{mfp,i}$ and the Larmor radius $r_{Li}$ are respectively $%
\lambda _{D}\simeq 10^{-2}-10^{6}cm$, $\lambda _{mfp,i}\simeq
10^{6}-10^{31}cm$ (an extremely large number!) and $r_{Li}\simeq
10^{-4}-10^{7}cm$ (the lower values corresponding to lower
temperature and higher density in the first and second cases and
to lower temperature and higher magnetic field in the third one).
For a plasma consisting of ion and electron species ($s=i,e$), one
can determine the range of values of these characteristic
parameters for each species. For any phenomena occurring on
timescales $\Delta t$ and lengthscales $\Delta L$, satisfying
\begin{eqnarray}
\tau _{ps},\tau _{Ls} &\ll &\Delta t\ll \tau _{Cs},  \label{time range} \\
\lambda _{D} &\ll &\Delta L\ll \lambda _{mfp,s},
\label{scale-length range}
\end{eqnarray}%
these plasmas can be considered as:

\begin{itemize}
\item (\#1) \emph{Collisionless:} due to the inequalities between
$\Delta t$ and $\tau _{Cs}$ and between $\Delta L$ and $\lambda
_{mfp,s}$. \emph{Any} plasma is effectively collisionless for
processes whose timescales and lengthscales are short enough and,
under those circumstances, one needs to use kinetic theory. For a
plasma which is sufficiently diffuse, this can include almost
\emph{all} relevant processes.

\item (\#2) Characterized by a \emph{mean-field EM interaction:}
due to the
inequalities between $\Delta t$ and $\tau _{ps}$ and between $\Delta L$ and $%
\lambda _{D}$, charged particles of the accretion disc interact
with the others only via a continuum mean EM field.

\item (\#3) \emph{Quasi-neutral:} due to the inequality between
$\Delta L$ and $\lambda_{D}$, the plasma can be taken as being
quasi-neutral on the lengthscale $\Delta L$.
\end{itemize}

When conditions \#1 and \#2 are satisfied, the medium is referred
to as a \emph{Vlasov-Maxwell plasma} and kinetic theory needs to
be used. It is this which we will be considering in the following.
The plasma is treated as an
ensemble of particle species, each being described by a KDF $f_{s}(\mathbf{y}%
,t)$ where $\mathbf{y}$ is the particle state vector and $t$ is
the time,
with $\mathbf{y}\equiv \left( \mathbf{r},\mathbf{v}\right) $ where $\mathbf{r%
}$ is the position and $\mathbf{v}$ is the velocity. The species
KDFs satisfy the Vlasov kinetic equation
$\frac{d}{dt}f_{s}(\mathbf{y},t)=0$, with the velocity moments of
the KDFs determining the source of the EM self-field $\left\{
\mathbf{E}^{self},\mathbf{B}^{self}\right\} $,
identified with the plasma charge and current density $\left\{ \rho (\mathbf{%
r},t),\mathbf{J}(\mathbf{r},t)\right\} $:
\begin{eqnarray}
\rho (\mathbf{r},t) &=&\sum\limits_{s}Z_{e}e\int
d^{3}vf_{s}(\mathbf{y},t),
\label{charge density} \\
\mathbf{J}(\mathbf{r},t) &=&\sum\limits_{s}Z_{e}e\int d^{3}v\mathbf{v}f_{s}(%
\mathbf{y},t).  \label{current density}
\end{eqnarray}%
We work here within the framework of perturbative kinetic theory,
extending the approach developed in Papers I and II. This approach
allows construction of analytical solutions for the KDF for
slowly-time-varying axisymmetric
gravitationally-bound systems (referred to as \textit{quasi-stationary KDFs}%
). The treatment is developed for both strongly and
weakly-magnetized plasmas, distinguishing between magnetic field
configurations with closed and open magnetic surfaces
(\cite{Coppi1,Coppi2,Cr2008-1,Cr2008-2}). In the present paper, we
are focusing on equilibrium or quasi-stationary configurations as
a preliminary to subsequently studying perturbations around these
solutions. Note that what is meant here by the term
\textquotedblleft equilibrium\textquotedblright\ is in general a
quasi-stationary KDF, expressed in terms of the relevant first
integrals of motion and adiabatic invariants (see Papers I and II
and \cite{Cr2008-1,Cr2008-2,Kocha10,Catto1987}), which
can also include a non-vanishing stationary radial accretion flow (\cite%
{Catania1}). One of our main results is the demonstration that for
strongly-magnetized collisionless plasmas, kinetic theory gives
the possibility of having quasi-stationary accretion flows even in
the absence of any additional forms of effective viscosity, such
as those arising from turbulence phenomena (which lie beyond the
equilibrium solutions). Within the framework of kinetic treatment,
quasi-stationary accretion flows can occur in collisionless AD
plasmas as part of the equilibrium configuration, with the role of
viscous stresses being played by the anisotropic pressure tensor
generated by phase-space anisotropies. We note that a somewhat
similar magnetically-driven non-turbulent mechanism for driving
accretion has been pointed out before (\cite{Blandford82}), based
on a purely fluid treatment. While that has some similarity with
what is being described here, in the sense that magnetic phenomena
in stationary configurations are giving rise to the redistribution
of angular momentum required for having the accretion flow, it is
very different in other respects.

We are concerned here particularly with investigating the role of
specifically kinetic effects which do not appear in a fluid
treatment. These can arise due to individual-particle dynamics
(e.g. finite Larmor-radius effects and ones arising from
microscopic conservation laws), as well as due to statistical
properties of the equilibrium KDF (e.g. temperature anisotropy and
non-uniform fluid fields). Their specific influence is found to
depend critically on the topology of the magnetic surfaces, the
statistical properties of the plasma and the strength of the EM
field. Instabilities and turbulent phenomena (see for example
\cite{Rebusco09,Lomi09}) are not considered here.

A number of issues arise for accretion-disc plasmas of this type,
including ones related to:

1) Existence of kinetic equilibria for both strongly and
weakly-magnetized plasmas;

2) Dynamo effects occurring in quiescent accretion-disc plasmas,
which can explain the self-generation of both azimuthal and
poloidal magnetic fields;

3) Collisionless non-turbulent quasi-stationary accretion
processes;

4) Closure conditions yielding a finite set of fluid equations for
the relevant fluid fields;

5) Equations of state for the pressure tensor components.

Investigation of these issues is relevant for correctly
understanding the equilibrium and dynamical properties of
collisionless accretion-disc plasmas.

\bigskip

The paper is organized as follows. In Section 2 we provide a
classification of accretion-disc plasmas together with the basic
assumptions and definitions being used. In Section 3, the first
integrals and adiabatic invariants of the system are derived, and
their physical meaning is discussed. Section 4 deals with the
construction of the quasi-stationary KDF for strongly-magnetized
plasmas with open magnetic surfaces. Similar calculations are then
presented in Section 5 for strongly-magnetized plasmas with closed
nested magnetic surfaces, and in Section 6 for weakly-magnetized
plasmas. In Section 7 the number density and flow velocity of each
species are computed for all of the configurations considered and
their physical properties are analyzed, stressing their connection
with accretion-disc dynamics. In Section 8 the pressure tensor for
each species is explicitly calculated. A separate discussion is
provided concerning the determination of equations of state for
the pressure tensor components. Section 9 deals with the
calculation of the fluid angular momentum conservation law, while
Section 10 deals with the kinetic accretion process which does not
depend on the presence of perturbative processes or an explicit
viscosity. Section 11 contains a demonstration of the existence of
a kinetic dynamo mechanism and presents equations giving the
poloidal and toroidal components of the magnetic field. Finally,
Section 12 contains a summary of the main results and closing
remarks. In the technical parts of this paper, we will often
abbreviate ``accretion-disc plasma'' to ``AD plasma'' for
conciseness.

\bigskip

\section{Basic physical assumptions}

In the following we will distinguish between strongly-magnetized,
intermediate and weakly-magnetized accretion-disc plasmas, the
distinction being made on the basis of asymptotic conditions
expressed in terms of suitable species-dependent dimensionless
physical parameters. These parameters will also be used for
characterizing the EM fields and for constructing perturbative
kinetic solutions for the quasi-stationary KDFs describing AD
plasmas. Three parameters are required:

1) The first parameter is $\varepsilon_{M,s}\equiv \frac{r_{Ls}}{L}$, where $%
s=i,e$ again denotes the species index. Here $r_{Ls}=v_{\perp
ths}/\Omega _{cs}$ is the species average Larmor radius, with
$v_{\perp ths}=\left\{ k_{B}T_{\perp s}/M_{s}\right\} ^{1/2}$
denoting the species thermal velocity perpendicular to the
magnetic field and $\Omega _{cs}=Z_{s}eB/M_{s}c$ being the species
Larmor frequency. Note, that we are here always measuring
temperatures in degrees Kelvin, with $k_{B}$ denoting the
Boltzmann constant. $L$ is the characteristic length-scale of the
spatial
inhomogeneities of the EM field, defined as $L\sim L_{B}\sim L_{E}$, where $%
L_{B}$ and $L_{E}$ are the characteristic magnitudes of the
gradients of the absolute values of the magnetic field
$\mathbf{B}\left( \mathbf{x},t\right)$
and the electric field $\mathbf{E}\left( \mathbf{x},t\right) $, defined as $%
\frac{1}{L_{B}}\equiv \max \left\{ \left\vert \frac{\partial }{\partial r_{i}%
}\ln B\right\vert ,i=1,3\right\} $ and $\frac{1}{L_{E}}\equiv \max
\left\{
\left\vert \frac{\partial }{\partial r_{i}}\ln E\right\vert ,i=1,3\right\} $%
, where the vector $\mathbf{x}$ denotes $\mathbf{x}=\left(
R,z\right) $.
From $\varepsilon _{M,s}$ it is possible to define a unique parameter $%
\varepsilon _{M}\equiv \max \left\{ \varepsilon
_{M,s},s=i,e\right\} $.

2) The second parameter is defined as $\varepsilon _{s}\equiv
\left\vert \frac{L_{\varphi s}}{p_{\varphi s}-L_{\varphi
s}}\right\vert =\left\vert \frac{M_{s}Rv_{\varphi
}}{\frac{Z_{s}e}{c}\psi }\right\vert $, i.e. it is the ratio
between the toroidal angular momentum of the particle $L_{\varphi
s}\equiv M_{s}Rv_{\varphi }$ and the magnetic contribution to its
toroidal
canonical momentum $p_{\varphi s}$, with $p_{\varphi s}=L_{\varphi s}+\frac{%
Z_{s}e}{c}\psi $. Here $v_{\varphi }\equiv \mathbf{v}\cdot \mathbf{e }%
_{\varphi }$, with $\mathbf{e}_{\varphi }$ being the unit vector
along the azimuthal direction $\varphi $, while $\psi $ denotes
the flux function of the poloidal magnetic field (see definition
below).

3) The third parameter is $\sigma _{s}\equiv \left\vert \frac{\frac{M_{s}}{2}%
v^{2}}{{Z_{s}e}\Phi _{s}^{eff}}\right\vert \simeq \left\vert \frac{k_{B}T_{s}%
}{{Z_{s}e}\Phi _{s}^{eff}}\right\vert $ which represents the ratio
between the kinetic energy of the particle $\frac{M_{s}}{2}v^{2}$
and its potential energy. The latter is represented in terms of
the effective potential $\Phi _{s}^{eff},$ which, in turn, is
related to the electrostatic and gravitational potentials (see the
next section for a rigorous definition), while for thermal plasmas
one has $\frac{M_{s}}{2}v^{2}\sim k_{B}T_{s}$. Related to $\sigma
_{s}$, we also define $\sigma \equiv \max \left( \sigma
_{s},s=i,e\right) $. The AD plasma is said to be
\emph{gravitationally bound} if $\sigma \ll 1$, and we are only
considering such cases in the following.

We classify these AD plasmas to be:

1) \emph{Strongly-magnetized} if the two conditions $\varepsilon
_{M,s}\ll 1$ and $\varepsilon _{s}\ll 1$ are satisfied.

2) \emph{Intermediately-magnetized }if $\varepsilon _{M,s}\ll 1$ but $%
\varepsilon _{s}\sim 1$.

3) \emph{Weakly-magnetized} if $\varepsilon _{M,s}\sim 1$ and
$\varepsilon _{s}\gtrsim 1$.

In both the strongly and intermediately magnetized cases, charged
particles spiral tightly around magnetic field lines (so that
gyrokinetic theory applies, see below), whereas they do not do
this in weakly-magnetized plasmas. In this sense, the behaviour of
strongly and intermediately magnetized plasmas is similar but, on
the other hand, they are importantly different in that the
magnetic flux $\psi$ completely dominates the particle canonical
momentum for strongly-magnetized plasmas, so that the contribution
due to the particle angular momentum $L_{\varphi s}$ is negligible
in the corresponding conservation law, whereas the contributions
are comparable for intermediately-magnetized plasmas.

To give an indication of the circumstances in which conditions 1)
and 2) above apply, we consider two situations representative of
stellar mass and galactic-centre mass black holes. Taking mass
$M_{*}=10M_{\odot}$ (giving a Schwarzschild radius $R_{Sch}\sim
30km$) as representative of a stellar-mass black hole, we focus on
AD plasma located at $R\sim 10\,R_{Sch}$ from the central object,
with ion temperature $T_{i}\sim 10^8K$, and characteristic
length-scale $L\sim 0.1R=R_{Sch}$ for the EM field. Then taking
$\varepsilon _{M,i}\lesssim 10^{-3}$ as representative of
$\varepsilon _{M,i}\ll 1$, we get $B\gtrsim 10\,G$ which is at the
lower end of the range of magnetic field strengths given above,
and so the condition $\varepsilon _{M,i}\ll 1$ can easily be
realized. For a galactic-centre black hole with mass
$M_{*}=10^8M_{\odot }$, the equivalent estimate, for the same
radial distance in terms of Schwarzschild radii, gives $B\gtrsim
10^{-6}\,G$. Considering now the condition $\varepsilon _{s}\ll
1$, we make similar choices as above, plus taking the
representative azimuthal particle velocity $v_{\varphi }$ as being
between the ion thermal velocity and the Keplerian velocity, and
the magnetic field strength as being related to the magnetic flux
$\psi$ by $B\sim\frac{\psi}{LR}$. Then requiring, for example,
$\varepsilon_{i}\lesssim 10^{-2}$ gives the (very approximate)
bounds of $B\gtrsim 10^{8}G$ for the stellar-mass case and
$B\gtrsim 10\,G$ for the galactic-centre one. The bound for the
stellar-mass case is at the upper end of the range for magnetic
field strengths given above, while that for the galactic-centre
one is far more modest (and we note that this is the more
interesting context in practice for the appearance of RIAF-type
accretion flows). However, concerning these estimates, one should
bear in mind that, in comparison with the case for $\varepsilon
_{M,s}$, the bound on $B$ coming from $\varepsilon _{s}$ is much
more sensitive to the species being considered (electrons or ions)
as well as to the value of the particle angular momentum and to
the approximate order-of-magnitude relationship used for
estimating $B$ from $\psi$. Depending on the circumstances being
considered, the condition $\varepsilon _{s}\ll 1$ may or may not
be satisfied, and so one should investigate both possibilities.

From the above considerations, it follows that both strongly and
intermediately-magnetized plasmas may be found in the inner
regions of accretion discs around compact objects.
Weakly-magnetized plasmas would be located further out, in the
outer regions of discs, where lower temperatures and weaker
magnetic fields are expected. We will focus here on studying the
configurations for strongly and weakly-magnetized collisionless AD
plasmas, leaving the corresponding investigation of
intermediately-magnetized plasmas to a separate study.

Ignoring possible weakly-dissipative effects (Coulomb collisions
and turbulence), we will assume that the KDF and the EM fields
associated with the plasma obey the system of Vlasov-Maxwell
equations, with Maxwell's equations being considered in the
quasi-static approximation. For definiteness, we will consider
here a plasma consisting of two species of charged particles: one
species of ions ($i$) and one of electrons ($e$).

Following the treatment presented in Papers I and II, we are
taking the AD plasma to be: a) \emph{non-relativistic}, in the
sense that it has non--relativistic species flow velocities and
$k_{B}T/m_{s}c^{2}$, that the gravitational field can be treated
within the classical Newtonian theory, and that the
non-relativistic Vlasov kinetic equation can be used as the
dynamical equation for the KDF;\ b) \emph{collisionless}, so that
the mean free path of the plasma particles is much longer than the
largest relevant characteristic scale length of the plasma; c)\
\emph{axisymmetric}, so that the relevant dynamical variables
characterizing the plasma (e.g., the fluid fields) are independent
of the azimuthal angle $\varphi ,$ when referred to a set of
cylindrical coordinates $(R,\varphi ,z)$; d) acted on by both
gravitational and EM fields.

We focus here on solutions for the equilibrium magnetic field
$\mathbf{B}$ which admit a family of locally-nested axisymmetric
toroidal magnetic surfaces (\cite{Coppi1,Coppi2}). We recall that
a magnetic surface is defined as a surface on which the poloidal
magnetic flux $\psi$ is constant, and that the condition $\nabla
\psi \cdot \mathbf{B}=0$ is then identically satisfied on each
magnetic surface. A schematic view of nested magnetic surfaces is
shown in Figure 1. For the equilibrium configurations which we are
considering here (prior to any subsequent perturbation) it is
reasonable to think that the magnetic field would be rather
ordered, at least on a local scale, so that having nested magnetic
surfaces is likely on that local scale. We distinguish between the
cases of \emph{locally closed} magnetic surfaces, discussed in
Paper I, and \emph{locally open} magnetic surfaces, discussed in
Paper II. Note that here the meaning of open and closed surfaces
has to be interpreted with reference to the local domain occupied
by the AD plasma. See Figure 1 for a schematic comparison of the
two topologies. For both of the configurations, a set of magnetic
coordinates ($\psi ,\varphi ,\vartheta $) can be defined locally,
where $\vartheta $ is a curvilinear angle-like coordinate on the
magnetic surfaces $\psi (\mathbf{x})=const.$ Each relevant
physical quantity $G(\mathbf{x},t)$ can then be conveniently
expressed either in terms of the cylindrical coordinates or as a
function of the magnetic coordinates, i.e.
$G(\mathbf{x},t)=\overline{G}\left( \psi ,\vartheta ,t\right) ,$
where the $\varphi $ dependence has been suppressed due to the
axisymmetry.

\begin{figure}
\centering
\includegraphics[width=3.5in,height=2.5in]{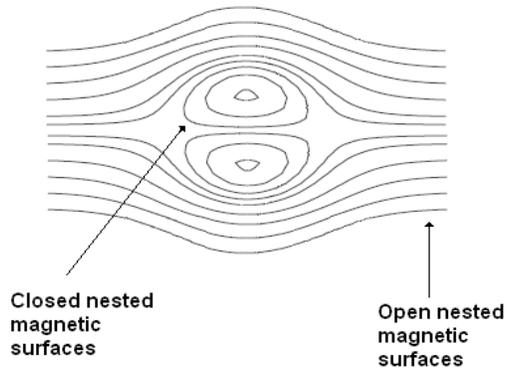} 
\caption{Schematic view of the topology of the magnetic surfaces.}
\end{figure}

For the configuration of closed nested magnetic surfaces, we will
assume \textquotedblleft small inverse aspect ratio
ordering\textquotedblright\ to hold, as expressed by the
requirement that $0<\delta \ll 1$. Here the dimensionless quantity
$\delta \equiv \frac{r_{\max }}{R_{0}}$ is referred to as the
\textit{inverse aspect ratio} parameter, where $R_{0}$ is the
radial distance from the vertical axis to the centre of the nested
magnetic surfaces and $r_{\max }$ is the average cross-sectional
poloidal radius of the largest closed toroidal magnetic surface.
This ordering is consistent with the results presented by
\cite{Coppi1,Coppi2,Cr2008-1, Cr2008-2} and with the assumption of
nested and closed magnetic surfaces which are assumed to be
localized in space (see also further discussion in Paper I).

In the following we will consider AD plasmas which are
characterized by slowly time-varying phenomena. In particular, the
EM field is taken to be given by an analytic function of the form
\begin{equation}
\left[ \mathbf{E}(\mathbf{x},\lambda
^{k}t),\mathbf{B}(\mathbf{x},\lambda ^{k}t)\right] ,  \label{b0}
\end{equation}%
with $k$ being an integer $\geq 1$ and $\lambda \equiv \min \left(
\varepsilon _{M},\sigma \right) $. This time dependence is
connected with either external sources or boundary conditions for
the KDF. In particular,
we will assume that the magnetic field is of the form%
\begin{equation}
\left. \mathbf{B}\equiv \nabla \times \mathbf{A}=\mathbf{B}^{self}(\mathbf{x}%
,\lambda ^{k}t)+\mathbf{B}^{ext}(\mathbf{x},\lambda ^{k}t),\right.
\label{b1}
\end{equation}%
where $\mathbf{B}^{self}$ and $\mathbf{B}^{ext}$ denote the
self-generated magnetic field produced by the AD plasma and a
finite external magnetic field produced by the central object (in
the case of neutron stars or white dwarfs). For greater
generality, we will not prescribe any relative orderings between
the various components of the total magnetic field, which are
taken to be of the form
\begin{eqnarray}
&&\left. \mathbf{B}^{self}=I(\mathbf{x},\lambda ^{k}t)\nabla
\varphi +\nabla \psi _{p}(\mathbf{x},\lambda ^{k}t)\times \nabla
\varphi ,\right.
\label{bself} \\
&&\left. \mathbf{B}^{ext}=\nabla \psi _{D}(\mathbf{x},\lambda
^{k}t)\times \nabla \varphi .\right.
\end{eqnarray}%
In particular, here $\mathbf{B}_{T}\equiv I(\mathbf{x},\lambda
^{k}t)\nabla \varphi $ and $\mathbf{B}_{P}\equiv \nabla \psi
_{p}(\mathbf{x},\lambda ^{k}t)\times \nabla \varphi $ are the
toroidal and poloidal components of the self-field, while the
external magnetic field $\mathbf{B}^{ext}$ is defined in terms of
the vacuum potential $\psi _{D}(\mathbf{x},\lambda ^{k}t) $. As a
consequence, the magnetic field can also be written in the
equivalent form
\begin{equation}
\mathbf{B}=I(\mathbf{x},\lambda ^{k}t)\nabla \varphi +\nabla \psi (\mathbf{x}%
,\lambda ^{k}t)\times \nabla \varphi ,  \label{B FIELD}
\end{equation}%
where the function $\psi (\mathbf{x},\lambda ^{k}t)$ is defined as $\psi (%
\mathbf{x},\lambda ^{k}t)\equiv \psi _{p}(\mathbf{x},\lambda
^{k}t)+\psi _{D}(\mathbf{x},\lambda ^{k}t),$ with $k\geq 1$ and
$(\psi ,\varphi ,\vartheta )$ defining a set of local magnetic
coordinates (as implied by the equation $\mathbf{B}\cdot \nabla
\psi =0$ which is identically satisfied). Finally, it is also
assumed that the charged particles of the
plasma are subject to the action of \emph{effective} \emph{EM potentials} $%
\left\{ \Phi _{s}^{eff}(\mathbf{x},\lambda ^{k}t),\mathbf{A}(\mathbf{x}%
,\lambda ^{k}t)\right\} ,$ where $\Phi
_{s}^{eff}(\mathbf{x},\lambda ^{k}t)$
is given by%
\begin{equation}
\Phi _{s}^{eff}(\mathbf{x},\lambda ^{k}t)=\Phi (\mathbf{x},\lambda ^{k}t)+%
\frac{M_{s}}{Z_{s}e}\Phi _{G}(\mathbf{x},\lambda ^{k}t),
\label{efp}
\end{equation}%
with $\Phi _{s}^{eff}(\mathbf{x},\lambda ^{k}t),$ $\Phi
(\mathbf{x},\lambda
^{k}t)$ and $\Phi _{G}(\mathbf{x},\lambda ^{k}t)$ denoting the \emph{%
effective} electrostatic potential and the electrostatic and
gravitational potentials. In particular, the gravitational
potential $\Phi _{G}$ is expressed as $\Phi _{G}=\Phi _{G}^{\left(
ext\right) }+\Phi _{G}^{\left( self\right) }$, where $\Phi
_{G}^{\left( ext\right) }$ and $\Phi _{G}^{\left( self\right) }$
denote the gravitational potentials due to the external sources
(i.e., primarily the central object) and the same accretion
disc\ respectively. The latter can be computed from the Poisson equation%
\begin{equation}
\nabla ^{2}\Phi _{G}^{\left( self\right) }(\mathbf{x},\lambda
^{k}t)=4\pi G_{N}\rho _{m}\left( \mathbf{x},\lambda ^{k}t\right) ,
\end{equation}%
where $G_{N}$ is the Newton gravitational constant and the source
term $\rho _{m}\left( \mathbf{x},\lambda ^{k}t\right) $ describes
the distribution of matter of the AD plasma. In the following it
is assumed that the gravitational potential $\Phi _{G}$ is a known
quantity.
\bigskip

\section{First integrals and adiabatic invariants for accretion-disc plasmas}

In this section we first define relevant kinetic conserved
quantities, namely dynamical variables depending on the states of
the individual charged particles of the plasma. We refer here in
particular to collisionless AD plasmas which satisfy the
assumptions introduced in the previous section. Then, on the basis
of this, we will investigate the basic physical implications of
the conservation laws for the qualitative properties of
single-particle dynamics. The determination of dynamical
invariants (i.e., first integrals of motion and adiabatic
invariants) is a basic requirement for the construction of
equilibrium solutions for the KDF (kinetic equilibria). We recall
here that an adiabatic invariant $P$ of order $n$ with respect to
$\lambda $ is a quantity which is conserved only
asymptotically, i.e., in the sense that $\frac{1}{\Omega _{cs}^{\prime }}%
\frac{d}{dt}\ln P=0+O(\lambda ^{n+1})$, where $n\geq 0$ is a
suitable integer. In other words, adiabatic invariants are
dynamical variables which change slowly on the time-scale of the
Larmor rotation time. These invariants can be derived from the
Lagrangian formulation of the single particle dynamics and the
corresponding gyrokinetic theory (see subsection below), which can
be obtained by means of a suitable asymptotic expansion in terms
of the dimensionless parameter $\varepsilon _{M,s}$ as long as
this is $\ll 1$. Because of this requirement, and in view of the
previous classification, gyrokinetic theory can only be formulated
for strongly-magnetized and intermediately-magnetized AD plasmas.

We first consider the treatment of the first integrals and
adiabatic invariants which are conserved for both
strongly-magnetized and weakly-magnetized AD plasmas. By
assumption, the only first integral of motion is the toroidal
canonical momentum $p_{\varphi s}$ conjugate to the
ignorable azimuthal angle $\varphi $:%
\begin{equation}
p_{\varphi s}=M_{s}R\mathbf{v\cdot e}_{\varphi
}+\frac{Z_{s}e}{c}\psi \equiv \frac{Z_{s}e}{c}\psi _{\ast s}.
\label{p_fi}
\end{equation}%
The total particle energy%
\begin{equation}
\left. E_{s}=\frac{M_{s}}{2}v^{2}\mathbf{+}{Z_{s}e}\Phi _{s}^{eff}(\mathbf{x}%
,\lambda ^{n}t),\right.  \label{total_energy}
\end{equation}%
is instead considered, by construction, to be an adiabatic
invariant of order $n$, with $n\geq 1$.

\subsection{Guiding-centre adiabatic invariants for strongly-magnetized
plasmas}

Additional adiabatic invariants can be determined for
strongly-magnetized plasmas based on gyrokinetic (GK) theory (see
Paper II, \cite{GK1,GK2,GK3}). We recall that gyrokinetic theory
provides a convenient formulation for single charged-particle
dynamics in the presence of EM fields which are strong enough so
that the motion of the particle can be well-represented as
spiraling around a single field line, following an imaginary
guiding centre which moves along the field line. Gyrokinetic
theory clearly displays the characteristic symmetry of the spiral
motion, allowing further conservation laws to be derived in
addition to those mentioned previously. We refer to Paper II for
the mathematical details concerning the formulation of
non-relativistic gyrokinetic theory in the presence of both EM and
gravitational fields. Here, for the sake of completeness, we
recall the basic notation for this and the main conclusions which
can be inferred from it. In the following, we will use a prime
\textquotedblleft\ $^{\prime }$ \textquotedblright\ to denote a
dynamical variable defined at the \emph{guiding-centre position} $\mathbf{r}%
^{\prime }$ (or $\mathbf{x}^{\prime }\equiv \left( R^{\prime
},z^{\prime }\right) $ in axisymmetry). The single particle
velocity is decomposed as
\begin{equation}
\mathbf{v}=u^{\prime }\mathbf{b}^{\prime }+\mathbf{w}^{\prime }+\mathbf{V}%
_{eff}^{\prime },
\end{equation}
where $u^{\prime }\mathbf{b}^{\prime }$ and $\mathbf{w}^{\prime }$
denote respectively the parallel and perpendicular components of
the guiding-centre velocity with respect to the magnetic field
direction, while the effective velocity $\mathbf{V}_{eff}^{\prime
}$ is defined as
\begin{equation}
\mathbf{V}_{eff}^{\prime }(\mathbf{x},\varepsilon _{M}^{k}t)\equiv \frac{c}{{%
B}^{\prime }}\mathbf{E}_{s}^{^{\prime }eff}\times
\mathbf{b}^{\prime }, \label{Vdrift}
\end{equation}
with $\mathbf{b}^{\prime
}=\mathbf{b}(\mathbf{x}^{\prime},\varepsilon
_{M}^{k}t),$ $\mathbf{b}(\mathbf{x},\varepsilon _{M}^{k}t)\mathbf{\equiv B}($%
\textbf{$\mathbf{x},$}$\varepsilon _{M}^{k}t)\mathbf{/}B(\mathbf{x}%
,\varepsilon _{M}^{k}t)$ being the unit vector of the local
magnetic field and where we have identified $\lambda =\varepsilon
_{M}$. It then follows
immediately that an adiabatic invariant is provided by the \textit{%
guiding-centre canonical momentum }$p_{\varphi s}^{\prime }$. Correct to $%
O(\varepsilon _{M}^{k})$, with $k\geq 1$, this is given by
\begin{equation}
p_{\varphi s}^{\prime }\equiv \frac{M_{s}}{B^{\prime }}\left(
u^{\prime }I^{\prime }-\frac{c\nabla ^{\prime }\psi ^{\prime
}\cdot \nabla ^{\prime }\Phi _{s}^{^{\prime }eff}}{B^{\prime
}}\right) +\frac{Z_{s}e}{c}\psi ^{\prime }.  \label{p_fi-HAT}
\end{equation}
A further adiabatic invariant is the \textit{magnetic moment}
$m_{s}^{\prime }$, which is proportional to the canonical momentum
$p_{\phi ^{\prime }s}$ conjugate to the gyrophase angle $\phi
^{\prime }$, which can be determined in principle to arbitrary
order in $\varepsilon _{M}$ (\cite{Kruskal}). In particular, the
leading-order approximation yields $m_{s}^{\prime }\cong \mu
_{s}^{\prime }\equiv \frac{M_{s}w^{\prime 2}}{2B^{\prime }}$.
Finally, by
construction, the \textit{guiding-centre Hamiltonian} $\mathcal{H}%
_{s}^{\prime }$ is also an adiabatic invariant. Accurate to order
$n$ this
is given by%
\begin{equation}
\mathcal{H}_{s}^{\prime (1)}\equiv m_{s}^{\prime }B^{\prime }+\frac{M_{s}}{2}%
\left( u^{\prime }\mathbf{b}^{\prime }+\mathbf{V}_{eff}^{\prime
}\right) ^{2}+{Z_{s}e}\Phi _{s}^{\prime eff}.  \label{HAM
giricinetica}
\end{equation}
When the above assumptions hold, the invariants determined in this
way necessarily exist for arbitrary initial conditions.

\subsection{Physical implications of the conservation laws}

We now discuss the physical meaning of the conservation laws
introduced here and their implications for particle dynamics in
magnetized accretion discs.

Consider first the conservation of the toroidal canonical momentum (\ref%
{p_fi}). For a charged particle it follows that this is the sum of
two terms: the particle angular momentum $L_{\varphi s}\equiv
M_{s}Rv_{\varphi }$ and a magnetic contribution
$\frac{Z_{s}e}{c}\psi $. It follows that the angular momentum by
itself is generally not conserved. As a consequence, the canonical
momentum conservation law allows for the existence of radial
particle motion inside a disc. In fact, since in AD plasmas the
magnetic flux function $\psi $ is necessarily spatially
non-homogeneous, a moving particle must change its angular
momentum $L_{\varphi s}$ while fulfilling
the constraint $\psi _{\ast s}=const.$, namely staying on a $\psi _{\ast s}-$%
surface. Depending on the geometry of the magnetic surfaces, such
particle motion may correspond to either a vertical or radial
velocity towards regions of higher or lower magnetic flux. Since
the magnetic contribution to $\psi _{\ast s}$ depends on the sign
of the charges, single ions and electrons exhibit motions in
different directions while keeping $\psi _{\ast s}$ constant. This
feature is different from the situation for neutral particles, for
which the angular momentum itself is conserved. Because of the
presence of plasma boundaries, this can lead to the
self-generation of quasi-stationary electric fields in the
accretion disc as a result of charge separation.

We next focus on the conservation of the guiding-centre Hamiltonian (\ref%
{HAM giricinetica}) and the magnetic moment $\mu _{s}^{\prime }$
(to leading-order approximation). These can be combined to
represent the
parallel velocity $u^{\prime }$ as%
\begin{equation}
u^{\prime }=\pm \sqrt{\frac{2}{M_{s}}\left[
\mathcal{H}_{s}^{\prime }-\mu
_{s}^{\prime }B^{\prime }-{Z_{s}e}\Phi _{s}^{\prime eff}-\frac{M_{s}}{2}%
V_{eff}^{\prime 2}\right] }.  \label{trap}
\end{equation}
Therefore $u^{\prime }$ is a local function of the guiding-centre
position vector $\mathbf{x}^{\prime }$ and, due to axisymmetry, of
the corresponding flux coordinates ($\psi ^{\prime },\vartheta
^{\prime }$). The above relationship is the basis of
\emph{particle trapping phenomena}, corresponding to the existence
of allowed and forbidden regions of
configuration space for the motion of charged particles. In fact, since $%
u^{\prime }$ is only defined in the subset of the configuration
space spanned by ($\psi ^{\prime },\vartheta ^{\prime }$) where
the argument of the square root is non-negative, it follows from
Eq.(\ref{trap}) that particles must undergo spatial reflections
when $u^{\prime }=0$. The points
of the configuration space where this occurs are the so-called \textit{%
mirror points} and the occurrence of such points may generate
various kinetic phenomena. In particular, particles can in
principle experience zero, one or two reflections corresponding
respectively to \textit{passing particles (PPs), bouncing
particles (BPs) }and \textit{trapped particles (TPs)}. In the
present case, since the right hand side of Eq.(\ref{trap}) depends
on the magnitude of the magnetic field ($B^{\prime }$), the
effective potential energy (${Z_{s}e}\Phi _{s}^{\prime eff}$) and
the centrifugal potential ($\frac{M_{s}}{2}V_{eff}^{\prime 2}$),
we will refer to the TP case as \emph{gravitational EM trapping}.

Finally, an important qualitative property of collisionless
magnetized plasmas follows from the conservation of the magnetic
moment $\mu _{s}^{\prime }$. The expression for this relates the
magnitude of the
perpendicular velocity $w^{\prime }$ to that of the local magnetic field $%
B^{\prime }$. Conservation of the adiabatic invariant $\mu
_{s}^{\prime }$ implies that when a charge is subject to a
non-uniform or a non-stationary
magnetic field, its kinetic energy of perpendicular motion, $M_{s}\frac{%
w^{\prime 2}}{2}$, must change accordingly so as to keep $\mu
_{s}^{\prime }$ constant. On the other hand, particles moving on
$\psi _{\ast s}-$surfaces
generally necessarily experience a non-uniform magnetic field $B(\mathbf{x}%
,\varepsilon _{M}^{k}t)$. It can be shown that this property
implies also the phenomenon of having a non-isotropic kinetic
temperature (i.e. there being different effective temperatures
parallel and perpendicular to the local direction of the magnetic
field). From the statistical point of view of kinetic theory, this
temperature anisotropy corresponds to an anisotropy in the kinetic
energy of random motion of particles subject to the magnetic
field. Such a feature is a characteristic kinetic phenomenon
arising in magnetized collisionless plasmas. This physical
mechanism operating at the level of single particle dynamics has
important consequences also for the macroscopic properties of such
plasmas. As we will see, conservation of $\mu _{s}^{\prime }$
allows the effects of temperature anisotropy to be included
consistently in the quasi-stationary solution for the KDF, and for
its physical implications for the dynamics of the corresponding
fluid system to be inferred. Another candidate source of
temperature anisotropy is radiation emission (cyclotron radiation)
due to Larmor rotation in the presence of a strong magnetic field.
The signature of this is the simultaneous occurrence of radiation
emission corresponding to the Larmor frequencies of the different
plasma species.

\section{The quasi-stationary KDF for strongly-magnetized plasmas: the case
of open surfaces}

In this section and the following one, we point out the most
relevant physical aspects of the kinetic treatment of
quasi-stationary strongly-magnetized collisionless AD plasmas in
the presence of magnetic field configurations with nested magnetic
surfaces. In this section, we consider the case of magnetic
surfaces which are open in the domain of the plasma. The case of
closed magnetic surfaces is treated in the following section.
Following the treatment presented in Paper II, here we want to
emphasize the physical aspects of the theory and the role of the
kinetic approach adopted here. For greater generality, it is
assumed that each species in the plasma is associated with a set
of \textit{sub-species} (the PPs, BPs and TPs mentioned above),
each one having a different KDF. The existence of temperature
anisotropy is allowed for all of the species, involving the
introduction of different temperatures parallel and perpendicular
to the local direction of the magnetic field. It is also assumed
that a non-vanishing species-dependent poloidal flow velocity can
exist, related to possible inward or outward matter flows in the
disc (see also the discussion below). Under these assumptions, as
pointed out in Papers I and II, a solution for the
quasi-stationary KDF can be obtained, which in the following is
denoted as $\widehat{f_{\ast s}}$. This is expressed in terms of
the integrals of motion and the adiabatic invariants identified in
the previous section. It has the form
\begin{equation}
\widehat{f_{\ast s}}=\widehat{f_{\ast s}}\left( E_{s},\psi _{\ast
s},p_{\varphi s}^{\prime },m_{s}^{\prime },\left( E_{s},\psi
_{\ast s}\right) ,\lambda ^{k}t\right) ,  \label{form}
\end{equation}%
where $k\geq 1$ and $\lambda $ is to be identified with
$\varepsilon _{M}$. Here, by construction, $\widehat{f_{\ast s}}$
is only defined in the subset of the phase-space where the
adiabatic invariants are defined, while the variables in the round
brackets $\left( E_{s},\psi _{\ast s}\right) $ will be involved in
the perturbative expansions to be defined below. It follows that
$\widehat{f_{\ast s}}$ is suitable for describing passing,
bouncing and trapped particles. Hence, by construction, the KDF is
itself an adiabatic invariant, and is therefore an asymptotic
solution of the Vlasov equation. Due to the arbitrariness of the
definition of the KDF, for each plasma sub-species it is always
possible to identify it with a superposition of Gaussian
distributions or, more generally, suitably generalized Gaussian
distributions. However, for a collisionless plasma, each of these
functions must actually itself be a quasi-stationary solution and
so, due to the requirement (\ref{form}), $\widehat{f_{\ast s}}$
can always be prescribed to be asymptotically \textquotedblleft
close\textquotedblright\ to a local
bi-Maxwellian KDF. Regarding this, in Paper II it was proved that $\widehat{%
f_{\ast s}}$ can be identified with a properly-defined
\textit{Generalized bi-Maxwellian KDF with parallel velocity
perturbations}.

In this paper we are mainly interested in the astrophysical
applications of the kinetic analysis and so we omit here all of
the mathematical details of the derivation, referring to Paper II
for an exhaustive discussion. It is sufficient to mention here
that a characteristic feature of the quasi-stationary KDF
$\widehat{f_{\ast s}}$ is that it contains implicit dependences in
terms of the single particle velocities. As shown in Paper II,
these dependences can be made explicit for strongly magnetized and
gravitationally bound plasmas so as to allow an asymptotic
analytical treatment of the velocity moments.

An interesting feature of the theory is the double
Taylor-expansion which is performed on $\widehat{f_{\ast s}}$ to
reach this goal, achieving a systematic solution method for the
Vlasov equation. More precisely, this is
done in terms of the two dimensionless parameters $\varepsilon _{s}$ and $%
\sigma _{s}$, requiring that the inequalities $\varepsilon _{s}\ll 1$ and $%
\sigma _{s}\ll 1$ are satisfied when the particle velocity is
taken to be of
the order $\left\vert \mathbf{v}\right\vert \lesssim v_{\perp ths}$ or $%
\left\vert \mathbf{v}\right\vert \lesssim v_{\parallel ths}$, where $%
v_{\perp ths}=\left\{ k_{B}T_{\perp s}/M_{s}\right\} ^{1/2}$ and $%
v_{\parallel ths}=\left\{ k_{B}T_{\parallel
s}/M_{s}\right\}^{1/2}$ denote respectively the parallel and
perpendicular thermal velocities. For greater generality,
$\varepsilon _{s}$ and $\sigma _{s}$ are here treated in the
perturbative expansions as being infinitesimals of the same order.
No expansion is performed in the ratio $\frac{\Omega
_{s}R}{v_{ths}}$. The Taylor expansion of $\widehat{f_{\ast s}}$
with respect to the two parameters can be formally carried out at
any order in $\varepsilon _{s}$
and $\sigma _{s}$. In particular, to leading-order this is done by setting $%
\psi _{\ast s}\cong \psi +O\left( \varepsilon _{s}\right) $ and $E_{s}\cong {%
Z_{s}e}\Phi _{s}^{eff}+O\left( \sigma _{s}\right) $. It is then
straightforward to prove that the following relation holds to
leading-order for the quasi-stationary KDF:
\begin{equation}
\widehat{f_{\ast s}}\cong \widehat{f_{s}}\left[ 1+h_{Ds}^{1}+h_{Ds}^{2}%
\right] ,  \label{solo}
\end{equation}%
where the leading-order distribution $\widehat{f_{s}}$ is of the form $%
\widehat{f_{s}}$ $=\widehat{f_{s}}\left( E_{s},\psi _{\ast
s},p_{\varphi s}^{\prime },m_{s}^{\prime },\left( \psi ,\vartheta
\right) ,\lambda ^{k}t\right) $. Here $h_{Ds}^{1}$ and
$h_{Ds}^{2}$ represent the so-called \emph{FLR-diamagnetic and
energy-correction parts} of $\widehat{f_{\ast s}}$
which are polynomial functions of the particle velocity, while $\widehat{%
f_{s}}$ can be written as%
\begin{eqnarray}
&&\left. \widehat{f_{s}}=\frac{n_{s}}{\left( 2\pi
k_{B}/M_{s}\right)
^{3/2}\left( T_{\parallel s}\right) ^{1/2}T_{\perp s}}\right.  \notag \\
&&\times \exp \left\{ -\frac{M_{s}\left( \mathbf{v}-\mathbf{V}%
_{s}-U_{\parallel s}^{\prime }\mathbf{b}^{\prime }\right) ^{2}}{%
2k_{B}T_{\parallel s}}-m_{s}^{\prime }\frac{B^{\prime }}{k_{B}\Delta _{T_{s}}%
}\right\} ,  \label{solo3}
\end{eqnarray}%
which we refer to as the \emph{bi-Maxwellian KDF with parallel
velocity perturbations}. Again we stress that $\widehat{f_{s}}$ is
only defined in the subset of phase-space where the parallel
velocity $\left\vert u^{\prime
}\right\vert $ is a real function. Here we note that the form of $\widehat{%
f_{s}}$ has been prescribed in order to allow the existence of:

1) A finite azimuthal flow velocity
$\mathbf{V}_{s}=\mathbf{e}_{\varphi }R\Omega _{s}$, with $\Omega
_{s}$ being a suitable rotational frequency (a toroidal angular
velocity).

2) Finite parallel flows (with respect to the local magnetic field
direction) associated with $U_{\parallel s}^{\prime }=\frac{I^{\prime }}{%
B^{\prime }}\xi _{s}$. These can include inward or outward radial
flows of matter. The guiding-centre parallel flow velocity
$U_{\parallel s}^{\prime }$ is uniquely prescribed in terms of
$\xi _{s}$, with $\xi _{s}$ being a suitable frequency (see Paper
II).

3) A finite toroidal magnetic field, which is related to
$U_{\parallel s}^{\prime }$. In fact $U_{\parallel s}^{\prime }$
is non-vanishing only for magnetic configurations in which the
toroidal magnetic field is non-zero.

4) A finite \emph{temperature anisotropy} identified with $\frac{1}{%
k_{B}\Delta _{T_{s}}}\equiv \frac{1}{k_{B}T_{\perp s}}-\frac{1}{%
k_{B}T_{\parallel s}}$.

5) A non-uniform effective number density defined as%
\begin{equation}
n_{s}=\eta _{s}\exp \left[ \frac{X_{s}}{k_{B}T_{\parallel
s}}\right] , \label{n1}
\end{equation}%
with $\eta _{s}$ denoting the \emph{pseudo-density}. Here the
function $X_{s} $ is prescribed in such a way as to take into
account the effects of the electrostatic and gravitational energy
of the particle, the centripetal potential and azimuthal and
parallel flows (the precise definition of this is given in Paper
II).

6) Separate treatments of species and sub-species contributions,
for which the previous asymptotic orderings are assumed to hold.
In fact, for the different populations the analytical expansion
can lead to different contributions for the terms appearing in the
diamagnetic and energy-correction parts, depending on the relative
magnitudes of the parameters $\varepsilon _{s}$ and $\sigma _{s}
$. On the other hand, because of the double expansion and the
energy dependence, the asymptotic solution for the two species can
hold also in different spatial domains.

Finally we note that the expansion given in Eq.(\ref{solo}) shows
that, in general, a bi-Maxwellian KDF cannot be an exact
stationary solution of the Vlasov equation. Instead, the actual
(asymptotic) equilibrium is necessarily described by the
quasi-stationary KDF $\widehat{f_{\ast s}}$. By construction this
is asymptotically close to a bi-Maxwellian when the expansion
(\ref{solo}) holds.

There are two important physical implications which follow from
the quasi-stationary solution for the KDF (\ref{solo}). The first
one concerns the existence of \emph{kinetic constraints}, namely
functional dependences which need to be imposed on the
quasi-stationary KDF in order to guarantee that this is an
adiabatic invariant of the prescribed order. To outline this
point, consider the set of functions
\begin{equation}
\Lambda _{s}\equiv \left( \beta _{s},\widehat{\alpha
_{s}},T_{\parallel s},\Omega _{s},\xi _{s}\right) ,
\end{equation}%
which we will refer to as \emph{structure functions} (see also
Paper II). In
particular, here%
\begin{eqnarray}
\beta _{s} &\equiv &\frac{\eta _{s}}{k_{B}T_{\perp s}}, \\
\widehat{\alpha _{s}} &\equiv &\frac{B^{\prime }}{k_{B}\Delta
_{T_{s}}},
\end{eqnarray}%
which depend on the pseudo-density, the magnitude of the
guiding-centre magnetic field and the parallel and perpendicular
temperatures. It is important to point out that the kinetic
constraints\ actually prescribe well-defined functional
dependences for the structure functions, imposing
for them the form%
\begin{equation}
\Lambda _{s}=\Lambda _{s}\left( \psi ,Z_{s}e\Phi _{s}^{eff}\right)
+O\left( \varepsilon _{s}\right) +O\left( \sigma _{s}\right) .
\label{kinkin}
\end{equation}%
The effective potential $\Phi _{s}^{eff}$ is generally a function
of the form $\Phi _{s}^{eff}=\Phi
_{s}^{eff}(\mathbf{x},\varepsilon _{M}^{k}t),$ with
$\mathbf{x}=\left( R,z\right) $, while neither the gravitational
potential nor the electrostatic potential are expected to be
functions only of $\psi $. Therefore, in magnetic coordinates, the
structure functions are of the general form $\Lambda _{s}\equiv
\overline{\Lambda }_{s}\left( \psi ,\vartheta ,\varepsilon
_{M}^{k}t\right) $. Hence, the functional forms of the
leading-order effective number density, the parallel and azimuthal
flow velocities and the temperatures carried by the bi-Maxwellian
KDF, are uniquely determined in terms of $\psi $ and $\vartheta $.
As a consequence, the azimuthal angular velocity is of the general
form $\Omega_{s} = \overline{\Omega }_{s}\left( \psi ,\vartheta
,\varepsilon _{M}^{k}t\right)$. We stress that in the customary
treatment of collisionless AD plasmas based on ideal-MHD these
constraints are missing. Instead, they follow in a natural way
from kinetic theory. By adopting a kinetic treatment it is
possible to prescribe the correct form for the fluid fields, as
required by the presence of kinetic constraints.

A further important consequence of the kinetic constraints is the
relationship between the magnitude of the temperature anisotropy
and the guiding-centre magnetic field at two different spatial
locations. In fact, the quantity $\frac{B^{\prime }}{\Delta
_{T_{s}}}$ in the KDF is necessarily an adiabatic invariant. To
leading-order in the GK expansion, this implies
that the asymptotic equation%
\begin{equation}
\frac{\left[ \Delta _{T_{s}}\right] _{2}}{\left[ \Delta _{T_{s}}\right] _{1}}%
\cong \frac{\left[ B\right] _{2}}{\left[ B\right] _{1}}
\end{equation}%
must hold identically for any two arbitrary positions
\textquotedblleft
1\textquotedblright\ and \textquotedblleft 2\textquotedblright , with $%
\left( \left[ \Delta _{T_{s}}\right] _{1},\left[ B\right] _{1}\right) $ and $%
\left( \left[ \Delta _{T_{s}}\right] _{2},\left[ B\right]
_{2}\right) $ denoting the temperature anisotropy and the
magnitude of the magnetic field at these positions respectively.

The second striking aspect of the kinetic treatment concerns the
diamagnetic
and energy-correction contributions $h_{Ds}^{1}$ and $h_{Ds}^{2}$ of $%
\widehat{f_{\ast s}}$. These carry the contributions from the
expansions of the particle toroidal canonical momentum and
particle total energy respectively. The perturbative-correction to
the KDF is a polynomial function of the particle velocity which
depends linearly on the so-called
effective \textit{thermodynamic forces. }The latter are here denoted as $%
A_{is}$ and $C_{is}$, with $i=1,5$. In analogy with classical
thermodynamics, it is natural to identify them with the gradients
of the structure functions\ $\Lambda _{s}$. Hence, in the present
case they are
associated with partial derivatives taken with respect to the magnetic flux $%
\psi $ and the effective potential $\Phi _{s}^{eff}$
(\cite{Cr2011}). There
are the following definitions: $A_{1s}\equiv \frac{\partial \ln \beta _{s}}{%
\partial \psi },$ $A_{2s}\equiv \frac{\partial \ln T_{\parallel s}}{\partial
\psi },$ $A_{3s}\equiv \frac{\partial \ln \Omega _{s}}{\partial \psi },$ $%
\widehat{A_{4s}}\equiv \frac{\partial \widehat{\alpha _{s}}}{\partial \psi }%
, $ $A_{5s}\equiv \frac{\partial \ln \xi _{s}}{\partial \psi }$ and $%
C_{1s}\equiv \frac{\partial \ln \beta _{s}}{\partial \Phi _{s}^{eff}},$ $%
C_{2s}\equiv \frac{\partial \ln T_{\parallel s}}{\partial \Phi
_{s}^{eff}},$
$C_{3s}\equiv \frac{\partial \ln \Omega _{s}}{\partial \Phi _{s}^{eff}}$, $%
\widehat{C_{4s}}\equiv \frac{\partial \widehat{\alpha
_{s}}}{\partial \Phi _{s}^{eff}},$ $C_{5s}\equiv \frac{\partial
\ln \xi _{s}}{\partial \Phi _{s}^{eff}}$.

The diamagnetic and energy-correction effects carried by $h_{Ds}^{1}$ and $%
h_{Ds}^{2} $ cannot be ignored: the construction of kinetic
equilibria cannot be achieved without them, as pointed out in
Papers I and II. From a physical point of view, the perturbative
contribution to the KDF determines first-order corrections to the
fluid moments of $\widehat{f_{\ast s}}$ produced by finite
Larmor-radius (FLR) effects. These carry the contributions of all
of the thermodynamic forces which can arise in collisionless
plasmas characterized by non-uniform differential rotation,
density and temperature gradients and temperature anisotropy.

\bigskip

\section{The quasi-stationary KDF for strongly-magnetized plasmas: the case
of closed surfaces}

In this section we present the kinetic solution for
strongly-magnetized plasmas in the case in which the equilibrium
magnetic field admits locally a family of closed and nested
magnetic surfaces. This is the configuration considered in Paper
I; the geometry is illustrated schematically in Fig.1.

The quasi-stationary KDF for collisionless plasmas with closed
magnetic surfaces can be found as a particular asymptotic limit of
the general solution $\widehat{f_{\ast s}}$ holding for open
surfaces. Besides considering closed surfaces, this is obtained by
imposing the requirement of small inverse aspect ratio ordering.
As a side assumption, we must now also impose vanishing of the
velocity perturbation $U_{\parallel s}^{\prime }$, which requires
setting $\xi _{s}$ to zero. In fact, the case of closed surfaces
corresponds to plasma magnetic self-confinement in which no local
net radial flow can take place, the latter being associated with $%
U_{\parallel s}^{\prime }$ (see Section 10). Note however that a
general situation can include both locally-closed and open
magnetic surfaces, as shown in Fig.1. In this context, the
quasi-stationary KDF $\widehat{f_{\ast s}}$ is reduced to a
\textit{Generalized bi-Maxwellian KDF} of the form
\begin{equation}
\widehat{f_{\ast s}}=\widehat{f_{\ast s}}\left( E_{s},\psi _{\ast
s},m_{s}^{\prime },\left( \psi _{\ast s}\right) ,\lambda
^{k}t\right) ,
\end{equation}
with $k\geq 1$ and $\lambda $ being identified with $\varepsilon
_{M}$, while the perturbative expansion is applied only to the
variable $\left( \psi _{\ast s}\right) $. When our set of
assumptions holds, the Taylor expansion of $\widehat{f_{\ast s}}$
can be performed only with respect to the dimensionless parameter
$\varepsilon _{s}$, while again no expansion is performed in the
ratio $\frac{\Omega _{s}R}{v_{ths}}$. Also, it is possible to
prove that in the present case the perturbative contributions with
respect to the expansion in $\sigma _{s}$ all become negligible.
More precisely, correct to first-order in $\varepsilon _{s}$, the
asymptotic expansion of $\widehat{f_{\ast s}}$ gives:
\begin{equation}
\widehat{f_{\ast s}}\cong \widehat{f_{s}}\left[
1+h_{Ds}^{1}\right] , \label{solo2}
\end{equation}%
with $\widehat{f_{s}}$ being of the form $\widehat{f_{s}}=\widehat{f_{s}}%
\left( E_{s},\psi _{\ast s},m_{s}^{\prime },\left( \psi \right)
,\lambda ^{k}t\right) $. The notation here is similar to that
adopted in the previous
section, with $h_{Ds}^{1}$ representing the \emph{diamagnetic part} of $%
\widehat{f_{\ast s}}$, while the leading-order distribution
$\widehat{f_{s}}$
is now given by%
\begin{eqnarray}
&&\left. \widehat{f_{s}}=\frac{n_{s}}{\left( 2\pi
k_{B}/M_{s}\right)
^{3/2}\left( T_{\parallel s}\right) ^{1/2}T_{\perp s}}\right.  \notag \\
&&\times \exp \left\{ -\frac{M_{s}\left(
\mathbf{v}-\mathbf{V}_{s}\right) ^{2}}{2k_{B}T_{\parallel
s}}-m_{s}^{\prime }\frac{B^{\prime }}{k_{B}\Delta
_{T_{s}}}\right\} ,  \label{solo5}
\end{eqnarray}%
which is referred to as the \emph{bi-Maxwellian KDF}. The form of $%
\widehat{f_{s}}$ has been prescribed in order to allow the
existence of:

1) A finite non-uniform azimuthal flow velocity $\mathbf{V}_{s}=\mathbf{e}%
_{\varphi }R\Omega _{s}$, where $\Omega _{s}$ is the rotational
frequency (the toroidal angular velocity).

2) A finite \emph{temperature anisotropy} characterized by $\frac{1}{%
k_{B}\Delta _{T_{s}}}\equiv \frac{1}{k_{B}T_{\perp s}}-\frac{1}{%
k_{B}T_{\parallel s}}$.

3) A finite toroidal magnetic field, related just to FLR and
diamagnetic effects driven by temperature anisotropy.

4) A non-uniform number density defined as%
\begin{equation}
n_{s}=\eta _{s}\exp \left[ \frac{X_{s}}{k_{B}T_{\parallel
s}}\right] , \label{n2}
\end{equation}
with $\eta _{s}$ again denoting the \emph{pseudo-density} and the function $%
X_{s}$ being prescribed in terms of the particle effective
electrostatic energy, the centripetal potential and the azimuthal
velocity (see the definition in Paper I).

5) Separate treatments of the species and sub-species
contributions, for which the previous asymptotic ordering is
assumed to hold.

Note that, unlike in the open-surface case, the energy-correction
contribution $h_{Ds}^{2}$ coming from the $\sigma _{s}$-expansion
does not appear in the asymptotic solution (\ref{solo2}), as this
becomes of higher order than $h_{Ds}^{1}$.

In this case the structure functions $\Lambda _{s}$ are
\begin{equation}
\Lambda _{s}\equiv \left( \beta _{s},\widehat{\alpha
_{s}},T_{\parallel s},\Omega _{s}\right) ,
\end{equation}%
whose physical meaning has been pointed out in the previous
section. As before, to leading order in the asymptotic expansion
(\ref{solo2}) and in the GK expansion, the structure functions
prescribe the fluid fields carried by the bi-Maxwellian KDF
(\ref{solo5}). In particular, for closed nested magnetic surfaces
and for strongly-magnetized plasmas, the kinetic
constraints give:%
\begin{equation}
\Lambda _{s}=\Lambda _{s}\left( \psi \right) +O\left( \varepsilon
_{s}\right) .  \label{lpsi}
\end{equation}
Comparison with Eq.(\ref{kinkin}) shows that the energy dependence
no longer appears, so that the fluid fields are only $\psi$-flux
functions. This is because in the present case, to leading order,
the effective potential is itself reduced to a flux-function, i.e.
$\Phi _{s}^{eff}=\Phi _{s}^{eff}(\psi ,\varepsilon_{M}^{k}t)$.
This conclusion is in agreement with the small inverse aspect
ratio ordering. Hence, the functional forms of the leading-order
number density, azimuthal flow velocities and temperatures
carried by the bi-Maxwellian KDF, are uniquely determined in terms of $%
\psi $.

Concerning the diamagnetic part $h_{Ds}^{1}$, it can easily be
shown that it carries the contributions arising from
Taylor-expanding the particle toroidal canonical momentum. As a
consequence, the diamagnetic KDF is a polynomial function of the
particle velocity, depending linearly on the thermodynamic forces
$A_{is}$, with $i=1,4$. From Eq.(\ref{lpsi}) and following the
treatments of Papers I and II, the latter are related to the
gradients of the structure functions with respect to the magnetic
flux $\psi
$, and are defined as follows: $A_{1s}\equiv \frac{\partial \ln \beta _{s}}{%
\partial \psi },$ $A_{2s}\equiv \frac{\partial \ln T_{\parallel s}}{\partial
\psi },$ $A_{3s}\equiv \frac{\partial \ln \Omega _{s}}{\partial \psi }$ and $%
\widehat{A_{4s}}\equiv \frac{\partial \widehat{\alpha _{s}}}{\partial \psi }$%
. The $A_{is}$, with $i=1,4$ carry the contributions due to
density, temperature, angular velocity and temperature anisotropy
gradients respectively. This general form of $h_{Ds}^{1}$\ follows
from the assumed kinetic equilibrium defined by Eq.(\ref{solo2}).
The diamagnetic effects carried by $h_{Ds}^{1}$ determine the
first-order FLR corrections to the fluid moments of
$\widehat{f_{\ast s}}$. They are important for characterizing
collisionless plasmas in the presence of non-uniform differential
rotation, density and temperature gradients and non-uniform
temperature anisotropy.

\bigskip

\section{The quasi-stationary KDF for weakly-magnetized plasmas}

In this section we derive a solution for the quasi-stationary KDF
describing weakly-magnetized collisionless AD plasmas. The basic
difference from the strongly-magnetized regime is that gyrokinetic
theory does not hold for weakly-magnetized plasmas because single
particles are effectively not magnetically confined (the Larmor
radius is of the same order as the characteristic equilibrium
scale-length of the plasma, or larger). Hence, contrary to the
case for strongly-magnetized plasmas, guiding-centre adiabatic
invariants can no longer be obtained. Here we retain, however, a
number of physical features relevant for modelling
weakly-magnetized plasmas:

1) Isotropic temperature: for all of the species it is assumed
that the temperature is isotropic.

2) Nested magnetic flux surfaces: the magnetic field is assumed to
allow quasi-stationary solutions with magnetic flux lines
belonging either to closed or open nested magnetic surfaces.

3) Azimuthal flow velocity: the plasma is characterized by having
a primarily differential azimuthal flow velocity, whose
leading-order expression is $\mathbf{V}_{\varphi s}\cong \Omega
_{s}R^{2}\nabla \varphi $.

4) Fluid fields: the collisionless plasma is characterized by
non-uniform fluid fields, defined in terms of velocity moments of
the quasi-stationary KDF.

5) Kinetic constraints: suitable functional dependences must be
imposed so as to ensure that the KDF is an asymptotic solution of
the collisionless Vlasov equation. For weakly-magnetized plasmas,
the kinetic constraints are found to differ from those considered
before for strongly-magnetized plasmas. They include, in
particular, constraints on the species angular frequency $\Omega
_{s}$.

6) Analytic form: the KDF is required to be a smooth analytic
function.

Given the requirements 1)-6), the solution for the KDF cannot, in
general, be a Maxwellian. However, it is possible to show that
they can all be satisfied by a suitably-generalized Maxwellian,
the new solution being expressed only in terms of the first
integral (\ref{p_fi}) and the adiabatic invariant
(\ref{total_energy}). For clarity of notation, in the following we
will label with $f_{ws}$ the quasi-stationary KDF for
weakly-magnetized plasmas. This has the general form
\begin{equation}
f_{ws}=f_{ws}\left( E_{s},p_{\varphi s},\left( E_{s}\right)
,\lambda ^{k}t\right) ,
\end{equation}%
with $k\geq 1$ and $\lambda $ being identified with $\sigma $. In
this case,
the perturbative expansion is carried out only with respect to the variable $%
\left( E_{s}\right) $. In agreement with the above requirements, a
particular solution for $f_{ws}$ is given by:
\begin{equation}
f_{ws}=\frac{\eta _{ws}}{\left( 2\pi k_{B}/M_{s}\right) ^{3/2}T_{ws}^{3/2}}%
\exp \left\{ -\frac{\left[ E_{s}-\Omega _{ws}p_{\varphi s}\right] }{%
k_{B}T_{ws}}\right\} ,  \label{f_w}
\end{equation}%
which will be referred to as the \emph{Generalized Maxwellian KDF
for weakly-magnetized plasmas}, with $E_{s}$ and $p_{\varphi s}$
being defined respectively by Eqs.(\ref{total_energy}) and
(\ref{p_fi}). In analogy with the previous treatment, we now
introduce the following \textit{structure
functions}:%
\begin{equation}
\Lambda _{ws}\equiv \left( \eta _{ws},T_{ws},\Omega _{ws}\right) .
\label{str_weak}
\end{equation}%
In order that Eq.(\ref{f_w}) defines an adiabatic invariant, the
following
functional dependences must be imposed on the structure functions:%
\begin{equation}
\Lambda _{ws}=\Lambda _{ws}\left( E_{s}\right) ,  \label{kincon_w}
\end{equation}%
which represent the kinetic constraints for weakly-magnetized
plasmas. Note that, for this configuration, only a dependence in
terms of the particle total energy is retained. Due to the
assumption of having a gravitationally-bound plasma, this is the
only physically admissible choice for the structure functions.

Using Eqs.(\ref{p_fi})-(\ref{total_energy}), an equivalent
representation
for $f_{ws} $ is provided by the expression%
\begin{equation}
f_{ws}=\frac{n_{ws}}{\left( 2\pi k_{B}/M_{s}\right)
^{3/2}T_{ws}^{3/2}}\exp
\left\{ -\frac{M_{s}\left( \mathbf{v}-\mathbf{V}_{ws}\right) ^{2}}{%
2k_{B}T_{ws}}\right\} .  \label{f_w_2}
\end{equation}%
Here $\mathbf{V}_{ws}\equiv \Omega _{ws}R^{2}\nabla \varphi $,
while the
function $n_{ws}$ is defined as%
\begin{equation}
n_{ws}\equiv \eta _{ws}\exp \left[
\frac{X_{ws}}{k_{B}T_{ws}}\right] , \label{n_w}
\end{equation}%
with%
\begin{equation}
X_{ws}\equiv M_{s}\frac{\left\vert \mathbf{V}_{ws}\right\vert ^{2}}{2}+\frac{%
Z_{s}e}{c}\psi \Omega _{ws}-Z_{s}e\Phi _{s}^{eff}.
\end{equation}%
We conclude that the quasi-stationary KDF is an adiabatic
invariant because of the kinetic constraints (\ref{kincon_w}).

\bigskip

\subsection{Analytical expansion of $f_{ws}$}

Imposing the kinetic constraints introduces an implicit velocity
dependence in the structure functions. This can be explicitly
dealt with by performing an asymptotic analytic expansion of
$f_{ws}$, similar to that made for strongly-magnetized plasmas. In
view of the form of Eq.(\ref{kincon_w}), a convenient Taylor
expansion for $f_{ws}$ can be obtained here in terms of
the parameter $\sigma _{s}$; there is no expansion in the ratio $\frac{%
\Omega _{s}R}{v_{ths}}$. For gravitationally bound plasmas, we can set $%
E_{s}\cong {Z_{s}e}\Phi _{s}^{eff}+O\left( \sigma _{s}\right) $ to
leading order. Correspondingly, the linear approximation for the
structure functions, obtained neglecting corrections of $O\left(
\sigma _{s}^{k}\right) $ with $k\geq 2$, is
\begin{equation}
\Lambda _{ws}\cong \Lambda _{s}+\left( E_{s}-Z_{s}e\Phi
_{s}^{eff}\right) \left[ \frac{\partial \Lambda _{ws}}{\partial
E_{s}}\right] _{E_{s}=Z_{s}e\Phi _{s}^{eff}},
\end{equation}%
where%
\begin{equation}
\Lambda _{s}\equiv \left[ \Lambda _{ws}\right] _{E_{s}=Z_{s}e\Phi
_{s}^{eff}}.  \label{leadw}
\end{equation}%
Then, the following relation for $f_{ws}$ holds, correct to first-order in $%
\sigma _{s}$:%
\begin{equation}
f_{ws}\cong f_{ws}^{0}\left[ 1+h_{ws}\right] ,  \label{fwf}
\end{equation}%
where $f_{ws}^{0}\left( E_{s},p_{\varphi s},\left( Z_{s}e\Phi
_{s}^{eff}\right) ,\lambda ^{k}t\right) $ is the leading-order solution and $%
h_{ws}$ represents the first-order perturbative contribution
coming from the Taylor expansion. The leading-order solution
$f_{ws}^{0}$ can be expressed as
\begin{equation}
f_{ws}^{0}=\frac{n_{s}}{\left( 2\pi k_{B}/M_{s}\right) ^{3/2}T_{s}^{3/2}}%
\exp \left\{ -\frac{M_{s}\left( \mathbf{v}-\mathbf{V}_{s}\right) ^{2}}{%
2k_{B}T_{s}}\right\} ,  \label{f}
\end{equation}%
which we will refer to as the \textit{drifted Maxwellian KDF for
weakly-magnetized plasmas} (i.e., a Maxwellian in the species
co-moving frame having velocity $\mathbf{V}_{s}$), and which depends on the number density $n_{s%
\text{,}}$ temperature $T_{s}$ and azimuthal flow velocity $\mathbf{V}%
_{s}\equiv \Omega _{s}R^{2}\nabla \varphi $, with $\Omega _{s}$
representing the leading-order azimuthal rotational frequency. The
leading-order number density is then defined as
\begin{equation}
n_{s}\equiv \eta _{s}\exp \left[ \frac{X_{s}}{k_{B}T_{s}}\right] ,
\label{3a}
\end{equation}%
with%
\begin{equation}
X_{s}\equiv M_{s}\frac{\left\vert \mathbf{V}_{s}\right\vert ^{2}}{2}+\frac{%
Z_{s}e}{c}\psi \Omega _{s}-Z_{s}e\Phi _{s}^{eff}.  \label{xx}
\end{equation}%
From Eq.(\ref{leadw}), to leading-order the structure functions
$\Lambda _{s}\equiv \left( \eta _{s},T_{s},\Omega _{s}\right) $
must satisfy the
following kinetic constraints:%
\begin{equation}
\Lambda _{s}=\Lambda _{s}\left( \Phi _{s}^{eff}\right) +O\left(
\sigma _{s}\right) .  \label{lam}
\end{equation}%
Finally, the first-order correction $h_{ws}$ is again a polynomial
function
of the particle velocity and has the form:%
\begin{equation}
h_{ws}\equiv h_{ws}^{1}v^{2}+h_{ws}^{2}v^{2}\left( \mathbf{v}\cdot \mathbf{e}%
_{\varphi }\right) ,  \label{hws}
\end{equation}%
where we have explicitly singled out the dependences on the
particle velocity. The contributions appearing in Eq.(\ref{hws})
are defined as
follows:%
\begin{eqnarray}
h_{ws}^{1} &\equiv &\frac{M_{s}}{2Z_{s}e}\left[
\begin{array}{c}
D_{1s}+\frac{Z_{s}e\psi \Omega _{s}}{ck_{B}T_{s}}D_{3s}+ \\
+\left[ \frac{Z_{s}e\left[ \Phi _{s}^{eff}-\frac{\psi \Omega
_{s}}{c}\right]
}{k_{B}T_{s}}-\frac{3}{2}\right] D_{2s}%
\end{array}%
\right] ,  \label{hw1} \\
h_{ws}^{2} &\equiv &\frac{M_{s}^{2}R\Omega
_{s}}{2Z_{s}ek_{B}T_{s}}\left[ D_{3s}-D_{2s}\right] ,  \label{hw3}
\end{eqnarray}%
where we have introduced the following definitions for the
\textit{energy gradients} of the structure functions:
$D_{1s}\equiv \frac{\partial \ln \eta
_{s}}{\partial \Phi _{s}^{eff}}$, $D_{2s}\equiv \frac{\partial \ln T_{s}}{%
\partial \Phi _{s}^{eff}}$, $D_{3s}\equiv \frac{\partial \ln \Omega _{s}}{%
\partial \Phi _{s}^{eff}}$. These quantities can again be interpreted as
generalized thermodynamic forces.

\subsection{Properties and discussion}

A number of comments should be made about the properties of the
kinetic solution obtained here and its physical meaning:

1) The existence of Eq.(\ref{f_w_2}) demonstrates that
quasi-stationary drifted Maxwellian kinetic solutions exist also
for weakly-magnetized collisionless AD plasmas.

2) Similarly to the case of strongly-magnetized plasmas, the
Maxwellian KDF obtained here is generally not an exact solution in
a strict sense. However, within the validity of the asymptotic
expansion (\ref{fwf}), this becomes an asymptotic equilibrium
solution.

3) The leading-order expressions for the number density,
temperature and azimuthal flow velocity appearing in the
Maxwellian KDF (\ref{f}), are found to be functions of $\Phi
_{s}^{eff}$.

4) The first-order perturbation $h_{ws}$ allows one to include
kinetic effects arising in weakly-magnetized plasmas. These
contributions are due to the thermodynamic forces $D_{is}$,
$i=1,3$, which carry information about the energy gradients of the
structure functions.

5) In this case the equilibrium is compatible only with an
azimuthal flow velocity, so that accretion flows can only occur as
a result of turbulence phenomena and cannot be described as
equilibrium solutions (see Sections 9 and 10).

6) The choice (\ref{kincon_w}) adopted here for the kinetic
constraints is an intrinsic feature of gravitationally bound
weakly-magnetized plasmas. For this plasma regime, the particle
dynamics is mainly determined by the gravitational potential and,
in principle, also by the electrostatic potential, as shown by
Eq.(\ref{lam}). In contrast, for strongly-magnetized plasmas with
closed surfaces, the dynamics is determined mainly by the magnetic
field and the structure functions are functions of the poloidal
magnetic flux $\psi $.

\bigskip

\section{Number density and flow velocity}

In this section we address the calculation of the relevant fluid
fields associated with the quasi-stationary KDF for both strongly
and
weakly-magnetized plasmas. By definition, given a distribution function $%
f_{s}$, a generic fluid field is expressed as an integral of the
distribution over the velocity space, having the form%
\begin{equation}
\int_{\Gamma _{u}}d^{3}vZ\left( \mathbf{x}\right) f_{s},
\end{equation}%
where $Z\left( \mathbf{x}\right) $ is an arbitrary velocity-weight
function and $\Gamma _{u}$ denotes the appropriate velocity space
for the integration. In this section we focus attention on
computing the species number density and flow velocity for plasmas
in the high and low magnetic
field regimes. Using $Z\left( \mathbf{x}\right) =1$ and $Z\left( \mathbf{x}%
\right) =\mathbf{v}$, one obtains:

a) \textit{species number density}%
\begin{equation}
n_{s}^{tot}\equiv \int_{\Gamma _{u}}d^{3}vf_{s};
\end{equation}

b) \textit{species flow velocity}%
\begin{equation}
\mathbf{V}_{s}^{tot}\equiv \frac{1}{n_{s}^{tot}}\int_{\Gamma _{u}} d^{3}v%
\mathbf{v}f_{s}.
\end{equation}%
Knowledge of these two fluid moments is required in order to write
the Poisson and Ampere equations for studying the self-generated
EM fields. A basic feature of the present calculation is that the
fluid fields are computed analytically in closed form by adopting
the asymptotic analytic
expansions of the quasi-stationary KDFs for $\widehat{f_{\ast s}}$ and $%
f_{ws}$. Note that these velocity moments are uniquely determined
once the quasi-stationary KDFs $\widehat{f_{\ast s}}$ and $f_{ws}$
are prescribed in terms of the structure functions $\Lambda _{s}$
and $\Lambda _{ws}$. On the other hand,\textit{\ the equilibrium
fluid moments which follow from this calculation are identically
solutions of the corresponding fluid moment equations.} These can
be obtained as velocity integrals of the Vlasov equation, of the form%
\begin{equation}
\int_{\Gamma _{u}}d^{3}vZ\left( \mathbf{x}\right)
\frac{d}{dt}f_{s}=0. \label{zzss}
\end{equation}
Finally, concerning the notation adopted, here and in the rest of
the paper the suffix ``tot'' is used to label fluid fields
expressed in terms of $\widehat{f_{\ast s}}$ or $f_{ws}$, and to
distinguish them from their leading-order solutions computed by
means of the asymptotic expansions of the same KDFs.
\subsection{Strongly-magnetized plasmas}

For strongly-magnetized plasmas, the fluid fields must be computed
by first performing a transformation of all of the guiding-centre
quantities
appearing in the quasi-stationary KDF to the actual particle position (%
\textit{guiding-centre back-transformation}). The order of
accuracy of this transformation is measured in terms of the
parameter $\varepsilon _{M,s}$ and depends on the corresponding
order of accuracy of the adiabatic invariants used in the solution
for the KDF. Contributions coming from this transformation carry
FLR corrections to the fluid fields, which operate together with
the FLR-diamagnetic and energy-correction contributions carried by
the first-order perturbations of the KDFs. Here a distinction must
be made between the cases of open and closed magnetic surfaces. In
fact, for open-field configurations it is possible to show that
$\varepsilon _{M,s}\lesssim \varepsilon _{s}$, while for
closed-field configurations,
within the validity of the inverse aspect ratio ordering, it follows that $%
\frac{\varepsilon _{M,s}}{\varepsilon _{s}}\sim O\left( \delta
\right) \ll 1. $ Hence, FLR effects from the guiding-centre
back-transformation are negligible for strongly-magnetized plasmas
with closed magnetic surfaces. Also, as indicated above, for
closed-field configurations the terms contributing to the KDF
coming from the $\sigma _{s}$-expansion are also negligible with
respect to those proportional to $\varepsilon _{s}$ and compared
with the open-field case. From these considerations it follows
that, when only first-order contributions are retained in the
asymptotic expansion, then the corresponding first-order
corrections coming from the guiding-centre back-transformation
need to be retained only for the leading-order KDF. Finally,
because of the existence of multiple-species plasmas, which for
strong magnetic fields\ may include also velocity-space
sub-species, the velocity sub-space $\Gamma _{u}$ of integration
must be properly prescribed. In fact, charged particles in both
open and closed configurations can have mirror points (TPs and
BPs) or be PPs, which are free to stream through the boundaries of
the domain. These populations give different contributions to the
relevant fluid fields and therefore require separate statistical
treatments.

We consider first the species number density. For the general case
of open magnetic surfaces the analytical expansion for the
quasi-stationary KDF is given by Eq.(\ref{solo}). Then, to
first-order in all of the expansion parameters ($\varepsilon
_{s}$, $\sigma _{s}$ and $\varepsilon _{M,s}$), the number density
is given by
\begin{equation}
n_{s}^{tot}\cong \int_{\Gamma _{u}} d^{3}v\left\{ \left[ \widehat{f_{s}}%
\right] _{GK}\left[ 1+h_{Ds}^{1}+h_{Ds}^{2}\right] \right\} ,
\label{a1}
\end{equation}%
where $\left[ \widehat{f_{s}}\right] _{GK}$ denotes the
leading-order KDF to which the guiding-centre back-transformation
must be applied up to first-order in $\varepsilon _{M,s}$. The
corresponding expression holding for closed surfaces in the case
of small inverse aspect ratio ordering
reduces to%
\begin{equation}
n_{s}^{tot}\cong \int_{\Gamma _{u}} d^{3}v\left\{
\widehat{f_{s}}\left[ 1+h_{Ds}^{1}\right] \right\} ,  \label{a2}
\end{equation}%
where we have made use of Eq.(\ref{solo2}). In both cases the
constitutive
equation for the total number density can be written as%
\begin{equation}
n_{s}^{tot}\cong n_{s}\left[ 1+\Delta _{n_{s}}\right] ,
\end{equation}%
where the leading-order contribution $n_{s}$ is given by
Eq.(\ref{n2}) for
closed surfaces and by Eq.(\ref{n1}) for open surfaces when the ordering $%
\varepsilon _{M,s}\ll \varepsilon _{s}$ holds, while $\Delta
_{n_{s}}$ carries the first-order corrections associated with FLR,
diamagnetic and energy-correction effects.

Next, let us consider the species flow velocity. Velocity-space
integrals analogous to (\ref{a1}) and (\ref{a2}) can be written
also in this case. The species flow velocity can be generally
represented in terms of the
constitutive equation%
\begin{equation}
n_{s}^{tot}\mathbf{V}_{s}^{tot}\cong n_{s}\left[ \mathbf{U}_{s}+\Delta _{%
\mathbf{U}_{s}}\right] .  \label{3}
\end{equation}%
For open magnetic surfaces and when $\varepsilon _{M,s}\ll \varepsilon _{s}$%
, $\mathbf{U}_{s}\equiv \mathbf{V}_{s}+U_{\parallel s}\mathbf{b}$
is the leading-order flow velocity carried by the bi-Maxwellian
KDF with a parallel velocity perturbation, in which
$\mathbf{V}_{s}=\Omega _{s}R^{2}\nabla \varphi $ and $U_{\parallel
s}=\frac{I}{B}\xi _{s}$. Moreover, the two frequencies $\Omega
_{s}$ and $\xi _{s}$ are subject to the kinetic
constraints given by Eq.(\ref{kinkin}). The second term $\Delta _{\mathbf{U}%
_{s}}$ represents the first-order correction which can be
decomposed as
follows:%
\begin{equation}
\Delta _{\mathbf{U}_{s}}=\left( \Delta _{\mathbf{U}_{s}\psi
}\nabla \vartheta \times \nabla \varphi ,\Delta
_{\mathbf{U}_{s}\varphi }\nabla \varphi ,\Delta
_{\mathbf{U}_{s}\vartheta }\nabla \psi \times \nabla \varphi
\right) .
\end{equation}%
Note that for open magnetic surfaces:

1) All components of $\Delta _{\mathbf{U}_{s}}$ are linear
functions of the thermodynamic forces appearing in the first-order
perturbation of the KDF.

2) The component $\Delta _{\mathbf{U}_{s}\varphi }$ provides a
correction to the leading-order azimuthal velocity. This is
non-vanishing even in the absence of any guiding-centre
contribution in the KDF and also in the case of isotropic
temperature.

3) The component $\Delta _{\mathbf{U}_{s}\vartheta }$ is related
to the temperature anisotropy as well as to both FLR-diamagnetic
and energy-correction effects.

4) The component $\Delta _{\mathbf{U}_{s}\psi }$ is associated
with FLR effects coming from the guiding-centre
back-transformation.

5) Both $U_{\parallel s}\mathbf{b}$ and $\Delta _{\mathbf{U}_{s}}$
can give rise to inward or outward flows of matter, both in the
radial and vertical directions. The existence of a non-vanishing
parallel velocity $U_{\parallel s}$ in the kinetic solution is
allowed by the gyrokinetic conservation laws. This means that it
can always be suitably prescribed in
agreement with the kinetic constraints. On the other hand, the correction $%
\Delta _{\mathbf{U}_{s}}$ acquires a precise physical meaning
within the present formulation. In fact $\Delta _{\mathbf{U}_{s}}$
is generated by the existence of a non-uniform and non-isotropic
plasma. Its precise form is determined automatically once the
kinetic constraints for the leading-order structure functions are
prescribed.

6) The expression for $\Delta _{\mathbf{U}_{s}}$ depends on the
particle sub-species (i.e., the distinction between PPs, TPs, and
BPs). In fact, each of these populations gives different
contributions to the components of the velocity $\Delta
_{\mathbf{U}_{s}}$. As indicated below, they give rise to
interesting physical phenomena for the disc dynamics, in
relationship with the Ampere equation and the self-generation of
both toroidal and poloidal equilibrium magnetic fields.

We consider now the case of closed magnetic surfaces. The
calculation is here made simpler because, by assumption,
$U_{\parallel s}$ vanishes and only the $\varepsilon
_{s}$-expansion is relevant in the asymptotic expansion. Thus, the
flow velocity can still be written as in Eq.(\ref{3}), but now
only an azimuthal leading-order flow velocity
$\mathbf{U}_{s}\equiv \Omega _{s}R^{2}\nabla \varphi $ can arise,
with $\Omega _{s}=\Omega _{s}\left( \psi \right) $ as indicated
above. On the other hand, the
first-order correction $\Delta _{\mathbf{U}_{s}}$ now reduces to%
\begin{equation}
\Delta _{\mathbf{U}_{s}}=\left( 0,\Delta _{\mathbf{U}_{s}\varphi
}\nabla \varphi ,\Delta _{\mathbf{U}_{s}\vartheta }\nabla \psi
\times \nabla \varphi \right) .  \label{v3}
\end{equation}%
In particular, the $\nabla \vartheta \times \nabla \varphi $
component vanishes in this approximation since it is related to
terms coming from the guiding-centre back transformation which are
of higher-order for closed surfaces. Again, the poloidal component
$\Delta _{\mathbf{U}_{s}\vartheta }$ is due to diamagnetic FLR
velocity corrections produced by temperature anisotropy (see also
Paper I). However, under the hypothesis of closed nested magnetic
surfaces, it cannot give rise to a net accretion velocity.

We therefore conclude that in both cases plasma temperature
anisotropy affects the existence of non-vanishing species poloidal
flow velocities. This feature only occurs for strongly-magnetized
plasmas. The importance of this result for AD plasmas lies in the
fact that equilibrium poloidal flow velocities may also give rise
to a net poloidal current density. The latter in turn will
generate a finite equilibrium toroidal magnetic field. Therefore,
species temperature anisotropies in collisionless AD plasmas
actually provide an effective physical mechanism for the
self-generation of toroidal magnetic field in these systems.

\subsection{Weakly-magnetized plasmas}

For weakly-magnetized plasmas, the explicit calculation of the
fluid fields is made simpler by the fact that the kinetic
equilibrium does not contain any guiding-centre adiabatic
invariant. In this case, only contributions arising from the
$\sigma _{s}$-expansion need to be taken into account. To
first-order in $\sigma _{s}$ the species number density can then
be easily
calculated as follows:%
\begin{equation}
n_{s}^{tot}\cong \int_{\Gamma _{u}}d^{3}v\left\{ f_{ws}^{0}\left[ 1+h_{ws}%
\right] \right\} ,
\end{equation}%
which recovers again the constitutive equation%
\begin{equation}
n_{s}^{tot}\cong n_{s}\left[ 1+\Delta _{n_{s}}\right] .
\end{equation}%
Here the leading-order term $n_{s}$ is defined by Eq.(\ref{3a})
and is the contribution carried by the Maxwellian KDF. On the
other hand, the
first-order correction $\Delta _{n_{s}}$ is found to be given by%
\begin{equation}
\Delta _{n_{s}}\equiv h_{ws}^{1}\left[ V_{s}^{2}+\frac{3k_{B}T_{s}}{M_{s}}%
\right] +h_{ws}^{2}\left[ \Omega _{s}^{2}R^{2}+\frac{9k_{B}T_{s}}{M_{s}}%
\right] \Omega _{s}R,
\end{equation}%
where $V_{s}^{2}\equiv \Omega _{s}^{2}R^{2}$ and the quantities
$h_{ws}^{1}$ and $h_{ws}^{2}$ are defined by
Eqs.(\ref{hw1})-(\ref{hw3}). Similarly, in
the same approximation, the flow velocity is given by the velocity integral%
\begin{equation}
\mathbf{V}_{s}^{tot}\cong \frac{1}{n_{s}^{tot}}\int_{\Gamma _{u}}d^{3}v%
\mathbf{v}\left\{ f_{ws}^{0}\left[ 1+h_{ws}\right] \right\} .
\end{equation}%
After explicit calculation, this gives%
\begin{equation}
n_{s}^{tot}\mathbf{V}_{s}^{tot}\cong n_{s}\left[ \mathbf{V}_{s}+\Delta _{%
\mathbf{V}_{s}}\right] ,  \label{v6}
\end{equation}%
where $\mathbf{V}_{s}\equiv \Omega _{s}R^{2}\nabla \varphi $ is
the leading-order flow velocity carried by the Maxwellian KDF. The
first-order
contribution $\Delta _{\mathbf{V}_{s}}$ is similarly found to be given by%
\begin{eqnarray}
\Delta _{\mathbf{V}_{s}} &\equiv &\mathbf{V}_{s}\Delta _{n_{s}}+\mathbf{V}%
_{s}h_{ws}^{1}\frac{6k_{B}T_{s}}{M_{s}}+  \notag \\
&&+ h_{ws}^{2}\left( \frac{15k_{B}T_{s}}{M_{s}}+V_{s}^{2}+2\Omega
_{s}^{2}R^{2}\right) \frac{k_{B}T_{s}}{M_{s}}\mathbf{e}_{\varphi
}.
\end{eqnarray}%
The following features should be noted:

1) The flow velocity\ is purely azimuthal. No further components
of the flow velocity are allowed for weakly-magnetized plasmas, in
contrast with the case of strongly-magnetized plasmas.

2) When the gravitational potential energy is dominant over the
electrostatic energy, the functional dependence of the structure
functions as implied by the kinetic constraints is determined
primarily by the gravitational potential. For example, the
leading-order azimuthal rotational frequency $\Omega _{s}$ is
simply of the form $\Omega _{s}\simeq \Omega _{s}\left( \Phi
_{G}\right) $. This means that its functional dependence is
compatible with the Keplerian rotational frequency.

\section{Pressure tensor and equations of state}

In this section we focus on the calculation of the pressure tensor
corresponding to the kinetic equilibria obtained for both strongly
and weakly-magnetized plasmas. The \textit{species pressure}
\textit{tensor }(or partial pressure tensor) is defined with
respect to the species
flow velocity as the velocity moment%
\begin{equation}
\underline{\underline{\Pi }}_{s}^{tot}\equiv \int_{\Gamma _{u}}
d^{3}vM_{s}\left( \mathbf{v}-\mathbf{V}_{s}^{tot}\right) \left( \mathbf{v}-%
\mathbf{V}_{s}^{tot}\right) f_{s}.  \label{tp}
\end{equation}
In the context of the present kinetic treatment $\underline{\underline{\Pi}}%
_{s}^{tot}$ is then uniquely prescribed in terms of the
quasi-stationary KDF. As indicated below, this enables us to
determine to the requisite accuracy also the corresponding
equations of state relating the components of the tensor to the
structure functions, giving a self-consistent treatment of the
physical properties of quasi-stationary collisionless AD plasmas.
In the
following we provide explicit expressions for $\underline{\underline{\Pi }}%
_{s}^{tot}$; the overall pressure tensor of the system can then be
obtained
by summing over the contributions from the separate species: $\underline{%
\underline{\Pi }}^{tot}=\sum_{s=i,e}\underline{\underline{\Pi
}}_{s}^{tot}$.

We consider first the case of strongly-magnetized plasmas,
adopting respectively for configurations with open and closed
magnetic surfaces the asymptotic expansions given by
Eqs.(\ref{solo}) and (\ref{solo2}). As a result of the
perturbative calculation, the species pressure tensor is
represented as%
\begin{equation}
\underline{\underline{\Pi }}_{s}^{tot}\simeq \underline{\underline{\Pi }}%
_{s}+\Delta \underline{\underline{\Pi }}_{s},  \label{ppppp}
\end{equation}%
where $\underline{\underline{\Pi }}_{s}$ is the leading-order term
(with
respect to all of the expansion parameters), while $\Delta \underline{%
\underline{\Pi }}_{s}$ represents the first-order correction. For
both
closed and open magnetic surfaces the tensor $\underline{\underline{\Pi }}%
_{s}$ is obtained using a bi-Maxwellian KDF with temperature
anisotropy. To represent $\underline{\underline{\Pi }}_{s}$, we
introduce for convenience
the set of right-handed orthogonal unit vectors $\left( \mathbf{b},\mathbf{e}%
_{1},\mathbf{e}_{2}\right) $, where $\mathbf{b}\equiv
\frac{\mathbf{B}}{B}$, while $\mathbf{e}_{1}$ and $\mathbf{e}_{2}$
are two orthogonal vectors in the plane perpendicular to the
magnetic field. In terms of this basis, the unit tensor can be
represented as: $\underline{\underline{\mathbf{I}}}\equiv
\mathbf{bb}+\mathbf{e}_{1}\mathbf{e}_{1}+\mathbf{e}_{2}\mathbf{e}_{2}$.
Then, it follows that with respect to the unit tensor $\underline{\underline{%
\mathbf{I}}}$, the pressure tensor $\underline{\underline{\Pi
}}_{s}$ is
symmetric, diagonal and non-isotropic, with a representation of the form:%
\begin{equation}
\underline{\underline{\Pi }}_{s}=n_{s}k_{B}T_{\perp s}\underline{\underline{%
\mathbf{I}}}+n_{s}k_{B}\left( T_{\parallel s}-T_{\perp s}\right)
\mathbf{bb}. \label{pres0}
\end{equation}%
Here $n_{s}$ is the leading-order species number density and
$T_{\parallel s} $, $T_{\perp s}$ are the leading-order species
parallel and perpendicular temperatures. Instead, the precise form
of $\Delta \underline{\underline{\Pi }}_{s}$ is geometry-dependent
and contains FLR corrections. It is always
possible to represent it in terms of the general decomposition:%
\begin{equation}
\Delta \underline{\underline{\Pi }}_{s}\equiv \Delta \Pi _{s}^{1}\underline{%
\underline{\mathbf{I}}}+\Delta \Pi _{s}^{2}\mathbf{bb}+\Delta \underline{%
\underline{\Pi }}_{s}^{3},  \label{deltap1}
\end{equation}%
in which $\Delta \Pi _{s}^{1}$ and $\Delta \Pi _{s}^{2}$ are
diagonal first-order anisotropic corrections to the pressure
tensor, while $\Delta \underline{\underline{\Pi }}_{s}^{3}$ in
this basis is generally non-diagonal. For strongly-magnetized
plasmas and closed magnetic surfaces, the precise form of the
tensor pressure has been given in \cite{Catania2}. The physical
properties of the solutions (\ref{pres0}) and (\ref{deltap1}) can
be summarized as follows:

1) The total tensor pressure $\underline{\underline{\Pi
}}_{s}^{tot}$ is symmetric.

2) The leading-order pressure tensor $\underline{\underline{\Pi
}}_{s}$ calculated in this approximation is diagonal but
non-isotropic. We note that the source of anisotropy in
Eq.(\ref{pres0}) is provided by the temperature anisotropy.

3) The first-order correction $\Delta \underline{\underline{\Pi
}}_{s}$
instead is generally non-diagonal and non-isotropic in the $\left( \mathbf{b}%
,\mathbf{e}_{1},\mathbf{e}_{2}\right) $ basis. Two different
physical mechanisms contribute to generating this effect. The
first one is again the temperature anisotropy, while the second is
produced by first-order perturbative corrections to the KDF, and
so depends linearly on the thermodynamic forces.

We next consider the case of weakly-magnetized plasmas. Performing
a similar calculation, it is possible to prove that, to
first-order in $\sigma _{s}$,
\textit{the pressure tensor is symmetric and isotropic} and can be written as%
\begin{equation}
\underline{\underline{\Pi }}_{s}^{tot}\simeq \left[ \Pi _{s}+\Delta \Pi _{s}%
\right] \underline{\underline{\mathbf{I}}}.  \label{pres4}
\end{equation}%
In particular, the leading-order term is defined as%
\begin{equation}
\Pi _{s}\equiv n_{s}k_{B}T_{s},
\end{equation}%
with the number density being given by Eq.(\ref{3a}). Both the
number density $n_{s}$ and the temperature $T_{s}$ are subject to
the kinetic constraints expressed by Eq.(\ref{lam}). The explicit
representation of the
first-order isotropic term $\Delta \Pi _{s}$ is as follows:%
\begin{eqnarray}
\Delta \Pi _{s} &\equiv &h_{ws}^{1}n_{s}\left[ \frac{15k_{B}^{2}T_{s}^{2}}{%
M_{s}^{2}}+V_{s}^{2}\frac{3k_{B}T_{s}}{M_{s}}\right] +  \notag \\
&&+h_{ws}^{2}n_{s}\Omega _{s}R\left[ \frac{45k_{B}^{2}T_{s}^{2}}{M_{s}^{2}}%
+V_{s}^{2}\frac{3k_{B}T_{s}}{M_{s}}\right] .  \label{zzzz}
\end{eqnarray}%
A striking feature of Eq.(\ref{zzzz}) is the explicit dependence
in terms of the thermodynamic forces, which take into account the
gradients of the
structure functions. Again, it should be noted that the pressure tensor (\ref%
{pres4}) is isotropic because the associated KDF has an isotropic
temperature and does not contain guiding-centre or FLR effects.

\subsection{Equilibrium equations of state}

An important result of the present theory is the explicit
construction of \textit{equations of state} for the various
components of the species pressure tensor. For definiteness, let
us consider here only the leading-order contributions to
$\underline{\underline{\Pi }}_{s}^{tot}$ with respect to the
relevant expansion parameters. This provides finite-term
equations (leading-order equations of state) of the form $\underline{%
\underline{\Pi }}_{s}^{tot}=\underline{\underline{\Pi
}}_{s}^{tot}\left( \Lambda _{s}\right) $, with $\Lambda _{s}$
being the appropriate structure
functions. Note that in principle the solution for $\underline{\underline{%
\Pi }}_{s}^{tot}$ allows one to obtain equations of state which
include also first-order corrections which are linearly
proportional to the thermodynamic forces.

In particular, in the case of strongly-magnetized plasmas this
recovers for the leading-order \textit{perpendicular and parallel
pressures} the
expressions%
\begin{eqnarray}
p_{\perp s} &\equiv &n_{s}k_{B}T_{\perp s},  \label{p_perp} \\
p_{\parallel s} &\equiv &n_{s}k_{B}T_{\parallel s}.  \label{p_par}
\end{eqnarray}
The number density and the parallel and perpendicular temperatures
are subject to their respective kinetic constraints (see the
discussion above). Eqs.(\ref{n1}) and (\ref{n2}) provide a clear
representation of the physical effects contributing to the
equations of state for strongly-magnetized plasmas in the case of
open and closed magnetic surfaces respectively. In particular, the
functions $X_{s}$ allow corrections due to particle electrostatic
and gravitational energy, centripetal potential and azimuthal and
parallel flows to be explicitly taken into account.

An analogous equation of state can be obtained in the case of
weakly-magnetized plasmas for the isotropic pressure tensor:%
\begin{equation}
p_{s}\equiv n_{s}k_{B}T_{s},  \label{p}
\end{equation}%
with $p_{s}$ denoting the species scalar pressure and $n_{s}$
being given by Eq.(\ref{3a}) to leading-order. This equation of
state allows one to clearly display the contributions due to the
gravitational and electrostatic potentials as well as the
azimuthal flow velocity.

Note that Eqs.(\ref{p_perp}) and (\ref{p_par}), as well as
Eq.(\ref{p}), provide only the leading-order solution for the
corresponding equations of state. In fact, a more accurate
solution should necessarily include also higher-order terms coming
from the diamagnetic and energy-correction contributions, both of
which result from the Taylor expansions performed on
the KDF. These corrections are responsible for the appearance of the terms $%
\Delta \underline{\underline{\Pi }}_{s}$ and $\Delta \Pi _{s}$ in Eqs.(\ref%
{ppppp}) and (\ref{pres4}) respectively. This implies that, for
both strongly and weakly magnetized collisionless plasmas, the
equations of state
for the partial pressures cannot be of the frequently-used polytropic type $%
p=\kappa \rho ^{\gamma }$ with $\kappa$ and $\gamma$ both being
constants (and $\rho$ again here being the mass density). This is
not surprising because the quasi-stationary kinetic solutions
considered here are not thermodynamic equilibria (as required for
deriving this polytropic equation of state from microscopic
considerations). This is clear from the appearance of
non-vanishing thermodynamic forces.

The present kinetic approach leads naturally to the use of
temperature and density as thermodynamic variables, since these
fluid fields are directly related to the (leading-order) structure
functions contained in the equilibrium KDF. On the other hand, for
a proton-electron plasma, which does not possess internal binding
energy, the temperature also represents a statistical measure of
the specific internal energy of the system, which by definition
does not include potential or EM energy. We note that equivalent
representations can be given in terms of temperature or of
specific internal energy although the more general case with a
non-isotropic pressure tensor requires the introduction of the
concept of \textquotedblleft directional\textquotedblright\
specific internal energy (related to the anisotropic
temperatures).

\section{Angular momentum}

In this section we discuss the implications of the kinetic
treatment for the law of conservation of fluid angular momentum.
For doing this, we first
define the \textit{species fluid canonical toroidal momentum} as%
\begin{equation}
L_{cs}^{tot}\equiv \frac{1}{n_{s}^{tot}}\int_{\Gamma _{u}}
d^{3}v\frac{Z_{s}e}{c}\psi _{\ast s}f_{s},
\end{equation}%
for $f_{s}=\widehat{f_{\ast s}}$ or $f_{s}=f_{ws}$ respectively in
the cases of strongly and weakly-magnetized plasmas. Consider then
the corresponding conservation law for the species total canonical
momentum. This can be
recovered by identifying the weight function $Z=\psi _{\ast s}$ in Eq.(\ref%
{zzss}), i.e. by setting%
\begin{equation}
\int_{\Gamma _{u}} d^{3}v\frac{d}{dt}\left[ \psi _{\ast
s}f_{s}\right] =0.
\end{equation}%
In the equilibrium case this implies the \textit{species fluid
angular momentum
conservation law}%
\begin{equation}
\nabla \cdot \left[ R^{2}\underline{\underline{\Pi
}}_{s}^{tot}\cdot \nabla \varphi
+n_{s}^{tot}\mathbf{V}_{s}^{tot}L_{s}^{tot}\right]
+\frac{Z_{s}e}{c}\nabla \psi \cdot
n_{s}^{tot}\mathbf{V}_{s}^{tot}=0  \label{angcons}
\end{equation}%
for the species angular momentum%
\begin{equation}
L_{s}^{tot}\equiv M_{s}R^{2}\mathbf{V}_{s}^{tot}\cdot \nabla
\varphi ,
\end{equation}%
where expressions for the number density, flow velocity and
pressure tensor have been derived in the previous sections. In
Eq.(\ref{angcons}) a key role is played by the divergence of the
species pressure tensor. For
strongly-magnetized plasmas, using the leading-order expression (\ref{pres0}%
), this is given by:%
\begin{equation}
\nabla \cdot \underline{\underline{\Pi }}_{s}^{tot}\cong \nabla p_{\perp s}+%
\mathbf{bB}\cdot \nabla \left( \frac{p_{\parallel s}-p_{\perp
s}}{B}\right) -\Delta p_{s}\mathbf{Q},
\end{equation}%
where $\mathbf{Q}\equiv \left[ \mathbf{bb}\cdot \nabla \ln B+\frac{4\pi }{cB}%
\mathbf{b}\times \mathbf{J}-\nabla \ln B\right] $ and $\Delta
p_{s}\equiv
\left( p_{\parallel s}-p_{\perp s}\right) $. It is clear that in this case $%
\nabla \cdot \underline{\underline{\Pi }}_{s}^{tot}$ has
non-vanishing components in arbitrary spatial directions,
including the azimuthal direction along $\nabla \varphi $. On the
other hand, for weakly-magnetized
plasmas, Eq.(\ref{pres4}) gives%
\begin{equation}
\nabla \cdot \underline{\underline{\Pi }}_{s}^{tot}\cong \nabla
\left[ \Pi _{s}+\Delta \Pi _{s}\right] \cdot
\underline{\underline{\mathbf{I}}}.
\end{equation}
Since the pressure tensor is isotropic and we are assuming
axisymmetry, it
follows that the component of $\nabla \cdot \underline{\underline{\Pi}}%
_{s}^{tot}$ along $\nabla \varphi$ must vanish identically.

For a single species, the total canonical momentum $L_{cs}^{tot}$
and the total angular momentum $L_{s}^{tot}$ in general differ
because of the contribution of the magnetic part proportional to
the flux function $\psi$. However, a different conclusion can be
drawn for the corresponding canonical momentum density
$n_{s}^{tot}L_{cs}^{tot}$ and angular momentum density
$n_{s}^{tot}L_{s}^{tot}$. If one considers summation over species
for both these quantities and imposes the \textit{quasi-neutrality
condition}
\begin{equation}
\sum\limits_{s}Z_{s}en_{s}^{tot}=0,
\end{equation}%
then one obtains the identity%
\begin{equation}
\sum\limits_{s}n_{s}^{tot}L_{s}^{tot}\equiv
\sum\limits_{s}n_{s}^{tot}L_{cs}^{tot}. \label{Ltot}
\end{equation}

We next investigate the consequences of Eq.(\ref{angcons}) for the
dynamical properties of collisionless plasmas. Note the following
aspects:

1) In the usual interpretation and also for weakly-magnetized
collisionless plasmas within our treatment (see Eq.(\ref{unaa})
below), the directional derivative of $L_{s}^{tot}$ along the flow
velocity $\mathbf{V}_{s}^{tot}$ vanishes. However, for
strongly-magnetized plasmas Eq.(\ref{angcons}) shows that
equilibrium configurations are possible in which this is generally
non-zero. This arises because of the non-isotropic pressure tensor
and the poloidal components of the flow velocity which, in turn,
are consequences of temperature anisotropy, the first-order
energy-correction and FLR-diamagnetic effects which are not
included in standard MHD treatments.

2) According to Eq.(\ref{angcons}), spatial variation in the
species angular momentum implies the possibility of having
quasi-stationary radial matter flows in the disc without departing
from the unperturbed equilibrium solution. These can correspond
either to local outflows or inflows; both can be described
consistently within the present kinetic solution for open magnetic
field lines in strongly-magnetized plasmas. Local inflows and
outflows can occur independently and are described consistently by
their respective quasi-stationary KDFs. Radial flows arise due
both to the parallel velocities $U_{\parallel s}$ and to the
kinetic effects driven by the first-order energy-correction and
FLR-diamagnetic effects. Therefore, species radial flows appear
necessarily together with a non-isotropic pressure tensor and a
non-vanishing toroidal magnetic field (see also Papers I and II
and the discussion below).

3) For weakly-magnetized plasmas, the tensor pressure is isotropic and Eq.(%
\ref{angcons}) reduces to%
\begin{equation}
n_{s}^{tot}\mathbf{V}_{s}^{tot}\cdot \nabla L_{s}^{tot} =0,
\label{unaa}
\end{equation}%
which, thanks to axisymmetry and recalling Eq.(\ref{v6}), is
identically satisfied. Hence, in this case equilibrium radial
flows are excluded.

In conclusion, for the case of weakly-magnetized plasmas, it
follows from the present treatment that quasi-stationary
equilibrium configurations with isotropic pressure tensor would
not have any net radial flow. This is as expected. In order to
have an accretion flow in this context one needs to have some form
of effective viscosity, either appearing explicitly or coming from
perturbations around the equilibrium state such as those leading
to MRI in the conventional picture. A particular goal of our
future work will be to investigate perturbations around the
equilibrium states presented here, to see whether a process
analogous to MRI then appears also for collisionless plasmas.

On the other hand, in the case of strongly-magnetized plasmas the
situation is different and net radial inflow can occur even in the
absence of effective viscosity being explicitly added or coming
from perturbative mechanisms. For strongly-magnetized plasmas,
Eq.(\ref{angcons}) implies that particles of one species can move
radially in a quasi-stationary configuration independently of
those of other species, with the species angular momentum not
being conserved, since the conservation law involves the canonical
momentum (including a magnetic-field contribution) and not just
the standard angular momentum. The flow velocity
$\mathbf{V}_{s}^{tot}$ is different for each species and, also,
the pressure tensor is non-isotropic. Hence, even with
quasi-neutrality, the total angular momentum of the matter can
change due to balance with the torque produced by the
non-isotropic species pressure tensor and/or a net current flow
across magnetic surfaces (see also
the discussion below). As indicated above, the non-isotropic nature of $%
\underline{\underline{\Pi } }_{s}^{tot}$ is caused by temperature
anisotropy as well as by first-order and FLR effects, which are
determined self-consistently in the kinetic approach.

\section{The kinetic accretion law}

In this section we discuss how quasi-stationary accretion flows
could occur in collisionless AD plasmas as a result just of the
equilibrium configuration without requiring additional effective
viscosity of the sort mentioned above. Here the role of viscous
stresses is played by the anisotropic pressure tensor, which is
part of the equilibrium solution.

For the physical conditions considered in the Introduction to
which the theory applies, the characteristic time for the inward
accretion flow in
accretion discs is typically longer than the characteristic Larmor time $%
\tau _{Ls}$ as well as the Langmuir time $\tau _{ps}$ and smaller
than the Spitzer ion collision time so that an accretion flow can
be consistently described by the present collisionless kinetic
treatment for quasi-stationary equilibria. In this section we will
demonstrate that equilibrium accretion flows cannot arise in the
case of strongly-magnetized plasmas with closed magnetic surfaces
and weakly-magnetized plasmas. Instead, equilibrium accretion
flows are permitted for general open-field configurations. Note,
however, that in general there are both open and closed surfaces
co-existing, as illustrated in Fig.1.

Let us consider first domains which are locally characterized by
open flux surfaces. In such domains, parallel flows can be
included in the
quasi-stationary KDF only if the guiding-centre canonical momentum $%
p_{\varphi s}^{\prime }$ is conserved, i.e. the plasma is
strongly-magnetized (see Section 4). In fact, in such cases, the
quasi-stationary KDF (\ref{solo}) can sustain both poloidal and
radial
species flow velocities, which are defined respectively as%
\begin{eqnarray}
V_{ps} &\equiv &\mathbf{V}_{s}^{tot}\cdot \mathbf{e}_{\vartheta }=\frac{1}{%
n_{s}^{tot}}\int_{\Gamma _{u}} d^{3}v\left[ \mathbf{v}\cdot \mathbf{e}%
_{\vartheta }\right] \widehat{f_{\ast s}}, \\
V_{Rs} &\equiv &\mathbf{V}_{s}^{tot}\cdot \mathbf{e}_{R}=\frac{1}{n_{s}^{tot}%
}J_{Rs}^{tot},
\end{eqnarray}%
where $\mathbf{e}_{\vartheta }\equiv \frac{\nabla \vartheta
}{\left\vert
\nabla \vartheta \right\vert },$ $\mathbf{e}_{R}\equiv \frac{\nabla R}{%
\left\vert \nabla R\right\vert }$ and the \textit{species mass
radial
current density }$J_{Rs}^{tot}$ is defined as%
\begin{equation}
J_{Rs}^{tot}\equiv \int_{\Gamma _{u}} d^{3}v\left[ \mathbf{v}\cdot \mathbf{e}%
_{R}\right] \widehat{f_{\ast s}}.
\end{equation}%
In the velocity-space integrals indicated above, the contributions
from PPs, BPs and TPs need to be distinguished.

The physically-relevant situations are those in which there is a
non-vanishing \textit{net radial species accretion flow}, i.e.
where the average species radial mass current $\left\langle
\left\langle J_{Rs}\right\rangle \right\rangle \equiv
\frac{1}{z_{2}-z_{1}}
\int_{z_{1}}^{z_{2}}J_{Rs}dz$ is negative (\textit{inward flow}), with $%
z_{1} $ and $z_{2}$ being suitably prescribed. Although local
contributions to $\left\langle \left\langle J_{Rs}\right\rangle
\right\rangle $ can arise from TPs, BPs and PPs, the overall
accretion flow is mainly associated with PPs. Notice also that, to
leading-order, the presence of poloidal accretion is necessarily
associated through the expression of $U_{\parallel s}$ to the
existence of a toroidal magnetic field. Note the following basic
features
involved in the accretion process. The ratio $\frac{\xi _{s}%
}{T_{\parallel s}}$ is approximately constant due to the kinetic
constraints, and so
\begin{equation}
\frac{\left[ \xi _{s}\right] _{1}}{\left[ \xi _{s}\right] _{2}}\cong \frac{%
\left[ T_{\parallel s}\right] _{1}}{\left[ T_{\parallel s}\right]
_{2}}
\end{equation}%
for any two arbitrary positions \textquotedblleft
1\textquotedblright\ and \textquotedblleft 2\textquotedblright\
prescribed in terms of the magnetic
coordinates $\left( \psi ,\vartheta \right) $. Then consider the case $%
\varepsilon _{M,s}\ll \varepsilon _{s}$, which allows one to
approximate the guiding-centre quantities with the expression for
them evaluated at the particle position. Assuming that $I=I\left(
\psi \right) $ (see Paper II)
and considering the two positions $\left( \psi ,\vartheta _{1}\right) $ and $%
\left( \psi ,\vartheta _{2}\right) $ on the same flux surface, the \textit{%
kinetic accretion law} follows%
\begin{equation}
\frac{U_{\parallel s}\left( \psi ,\vartheta _{1}\right)
}{U_{\parallel s}\left( \psi ,\vartheta _{2}\right) }\cong
\frac{B\left( \psi ,\vartheta _{2}\right) }{B\left( \psi
,\vartheta _{1}\right) }\frac{T_{\parallel s}\left( \psi
,\vartheta _{1}\right) }{T_{\parallel s}\left( \psi ,\vartheta
_{2}\right) }.  \label{aclaw}
\end{equation}%
In this case, under the same assumptions, it follows from the
continuity equation that the ratio of the corresponding species
number densities must
vary on a given $\psi $-surface according to the following relation:%
\begin{equation}
\frac{n_{s}\left( \psi ,\vartheta _{1}\right) }{n_{s}\left( \psi
,\vartheta
_{2}\right) }\cong \frac{B^{2}\left( \psi ,\vartheta _{1}\right) }{%
B^{2}\left( \psi ,\vartheta _{2}\right) }\frac{T_{\parallel
s}\left( \psi ,\vartheta _{2}\right) }{T_{\parallel s}\left( \psi
,\vartheta _{1}\right) }.
\end{equation}%
Therefore, on a given $\psi $-surface:

1) the species parallel flow velocity increases with the parallel
temperature while decreasing with respect to the magnitude of the
magnetic field;

2) the species number density instead increases with the magnetic
pressure and decreases with the parallel temperature.

The physical interpretation for both of these is clear: higher
magnetic pressure slows down the matter accretion rate while
increasing the number density, whereas higher parallel temperature
corresponds to higher radial fluid mobility, thus decreasing the
local species number density.

Next, we consider domains of strongly-magnetized plasmas with
closed field lines, again using the inverse aspect ratio
expansion. We want to prove that the poloidal-averaged radial flow
velocity $\left\langle V_{Rs}\right\rangle _{\vartheta }$ vanishes
identically, when $V_{Rs}$ is computed in terms of the
quasi-stationary KDF (\ref{solo2}). Given a generic function of
the form $C\left( \psi ,\vartheta \right) $, the operator
$\left\langle C\left( \psi
,\vartheta \right) \right\rangle _{\vartheta }$ is defined as%
\begin{equation}
\left\langle C\left( \psi ,\vartheta \right) \right\rangle
_{\vartheta }\equiv \frac{1}{\kappa }\int_{0}^{2\pi
}\frac{d\vartheta }{\left\vert \mathbf{B}_{p}\cdot \nabla
\vartheta \right\vert }C\left( \psi ,\vartheta \right) ,
\end{equation}%
with $\kappa \equiv \int_{0}^{2\pi }\frac{d\vartheta }{\left\vert \mathbf{B}%
_{p}\cdot \nabla \vartheta \right\vert }$ and so, from
Eq.(\ref{v3}) it
follows that, to leading-order in the inverse aspect ratio%
\begin{equation}
\left\langle V_{Rs}\right\rangle _{\vartheta }\cong \frac{1}{\kappa }%
\int_{0}^{2\pi }d\vartheta \frac{\mathbf{B}_{p}\cdot \nabla
R}{\left\vert
\mathbf{B}_{p}\cdot \nabla \vartheta \right\vert }\frac{\Delta _{\mathbf{U}%
_{s}\vartheta }\left( \psi \right) }{\left\vert \nabla
R\right\vert }\cong 0
\end{equation}%
which vanishes because, to leading-order in $\delta $, $\mathbf{B%
}_{p}\cdot \nabla R\cong B_{p}\nabla \vartheta \cdot \nabla R$ is
antisymmetric with respect to the transformation $\vartheta
\rightarrow \vartheta +\pi $. Higher-order corrections in the
inverse aspect ratio can be included, but they require also
performing the asymptotic expansion of the KDF to higher-order in
both $\sigma _{s}$ and $\varepsilon _{s}$. Therefore, in
sub-domains of the plasma where magnetic surfaces are closed and
nested no net equilibrium accretion flow can arise.

Finally, the corresponding treatment for weakly-magnetized plasmas
can be
recovered from Eq.(\ref{fwf}). Since the species flow velocity $\mathbf{V}%
_{s}^{tot}$ has only an azimuthal component, it follows that
$V_{Rs}$ is identically zero. Hence, under the present
assumptions, no net quasi-stationary radial flow can arise in the
case of weakly-magnetized plasmas.

To summarize: the present theory provides a possible new
collisionless physical mechanism giving an equilibrium accretion
process in AD plasmas. In particular, we note that:

1) Only strongly-magnetized plasmas with open magnetic surfaces
can sustain these equilibrium accretion flows.

2) The primary source of this equilibrium accretion flow mechanism
is the appearance of equilibrium radial flows driven by
temperature anisotropies and phase-space anisotropies. These are
directly connected with the existence of non-isotropic species
pressure tensors, which in turn play the role of an effective
viscosity in driving quasi-stationary accretion flows.

3) Quasi-stationary accretion flows are consistent with the basic
conservation laws (for mass density and canonical momentum) and
with the existence of a non-isotropic species pressure tensor (see
also the discussion in Section 11).

4) The accretion law could, in principle, be tested experimentally
if one had suitable observations, since it relates the magnitude
of the species parallel flow velocity $U_{\parallel s}$ to the
local values of the magnetic field magnitude and the parallel
temperature.

5) First-order (as well as higher-order) perturbative corrections,
can in principle be included consistently in the present theory.

\section{The kinetic dynamo}

Here we address the problem of the self-generation of magnetic
field in quasi-stationary collisionless AD plasmas. We refer here
to this phenomenon of
self-generation of both poloidal and toroidal magnetic fields as a \textit{%
quasi-stationary kinetic dynamo effect}. This is described by the
Ampere
equation with current density $\mathbf{J}$ defined as%
\begin{equation}
\mathbf{J}\equiv \sum\limits_{s}\mathbf{J}_{s}=\sum\limits_{s}Z_{s}e\int_{%
\Gamma _{u}} d^{3}v\mathbf{v}f_{s}
\end{equation}%
for $f_{s}=\widehat{f_{\ast s}}$ or $f_{s}=f_{ws}$ respectively in
the cases of strongly and weakly-magnetized plasmas. In all cases
considered here, the current density can be generally represented
in terms of the magnetic
coordinates $(\psi ,\varphi ,\vartheta )$ as%
\begin{equation}
\mathbf{J}=\left( J_{\psi }\nabla \vartheta \times \nabla \varphi
,J_{\varphi }\nabla \varphi ,J_{\vartheta }\nabla \psi \times
\nabla \varphi \right) .
\end{equation}%
In particular, based on the calculation of the flow velocity (see
Section 10), it can be shown that:

a) For strongly-magnetized plasmas with open magnetic surfaces all
of the three components $\left( J_{\psi },J_{\varphi
},J_{\vartheta }\right) $ are generally non-vanishing. These
include the contributions carried by TPs, BPs and PPs.

b) For strongly-magnetized plasmas with closed magnetic surfaces,
$J_{\psi }$ vanishes identically. In this case the current
includes contributions from both TPs and PPs.

c) For weakly-magnetized plasmas both $J_{\psi }$ and
$J_{\vartheta }$ vanish identically.

The toroidal component of the Ampere equation gives the
generalized
Grad-Shafranov equation for the poloidal flux function $\psi _{p}$:%
\begin{equation}
\Delta ^{\ast }\psi _{p}=-\frac{4\pi }{c}J_{\varphi },
\label{GSapp}
\end{equation}%
where $\Delta ^{\ast }\equiv R^{2}\nabla \cdot \left( R^{-2}\nabla
\right) $ is an elliptic differential operator. The components of
Ampere's equation along the directions $\nabla \vartheta \times
\nabla \varphi $ and $\nabla \psi \times \nabla \varphi $ provide
instead two PDEs for the toroidal component of the magnetic field
$\left\vert \mathbf{B}_{T}\right\vert =I/R$
(see the definition in Eq.(\ref{bself})):%
\begin{eqnarray}
\frac{\partial I}{\partial \psi } &=&\frac{4\pi }{c}J_{\vartheta }, \\
\frac{\partial I}{\partial \vartheta } &=&\frac{4\pi }{c}J_{\psi
},
\end{eqnarray}%
which give the general solubility constraint%
\begin{equation}
\frac{\partial J_{\psi }}{\partial \psi }=\frac{\partial J_{\vartheta }}{%
\partial \vartheta },  \label{solubility}
\end{equation}%
which is equivalent to imposing the charge continuity equation
$\nabla \cdot \mathbf{J}=0$. In the case of closed surfaces, the
additional solubility
condition%
\begin{equation}
\oint J_{\psi }d\vartheta =0  \label{period}
\end{equation}%
must also be imposed. Both constraint equations must be taken as
being solubility conditions to be satisfied due to the
arbitrariness of the structure functions.

For open surfaces, the function $I$ is of the general form $I(\psi
,\vartheta )$ and so Eq.(\ref{solubility}) can always be
satisfied. Instead, for closed surfaces $I=I\left( \psi \right) $,
which can always be asymptotically satisfied if inverse aspect
ratio ordering applies. Constraints (\ref{solubility}) and
(\ref{period}) are then both satisfied.

Various scenarios can be envisaged in which quasi-stationary
kinetic dynamos can be present. We will list the basic features of
this mechanism:

1) In all of the cases discussed above, $J_{\varphi }$ is
generally non-vanishing, implying the existence of a
self-generated poloidal magnetic field. Instead, a toroidal
component of the magnetic field only arises if a poloidal current
is present.

2) For strongly-magnetized plasmas, first-order FLR-diamagnetic
and energy-correction effects, driven by temperature anisotropy,
are responsible for the generation of poloidal currents, and hence
of toroidal magnetic field. Gyrophase-dependent contributions can
arise in this case, driven by the same thermodynamic forces. These
originate from the guiding-centre back-transformation, which is
characteristic of open-field configurations, and are responsible
for the generation of $J_{\psi }$. For strongly-magnetized plasmas
with open magnetic surfaces, the parallel velocity $U_{\parallel
s}$ also contributes to the generation of poloidal currents.

3) For weakly-magnetized plasmas, only the poloidal component of
the magnetic field can be self-generated. However, when the
azimuthal angular velocity coincides to leading-order with the
Keplerian frequency, the azimuthal current density vanishes due to
quasi-neutrality (to leading-order). Therefore, in this case the
current is necessarily produced by first-order corrections. This
conclusion is consistent with the assumption of weak magnetic
field.

We stress that, in contrast to customary MHD treatments, the
quasi-stationary kinetic dynamo effect described here can occur
even \textit{in the absence of possible instabilities or
turbulence phenomena}. In particular, configurations with closed
magnetic surfaces or contributions from TPs in the case of open
surfaces could be responsible for the self-generation of toroidal
field even \textit{without needing any net accretion flow} in the
domain of interest. This toroidal field is associated with the
existence of kinetic torques which cause redistribution of angular
momentum, as discussed above.

\section{Conclusions}

In this paper a kinetic description for collisionless
quasi-stationary accretion disc plasmas has been formulated within
the framework of Vlasov-Maxwell theory. A perturbative approach
has been developed, which enables the systematic analytical
construction of equilibrium kinetic distribution functions for
non-relativistic axisymmetric collisionless plasmas subject to
both gravitational and electromagnetic fields. The cases of both
weakly and strongly-magnetized plasmas have been investigated, for
configurations with both open and closed magnetic surfaces. The
main aim was to establish kinetic equilibrium configurations to
use as a starting point for subsequent perturbation analysis,
looking for kinetic mechanisms which could give rise to an
effective viscosity able to drive accretion flows within the
collisionless regime.

Equilibrium solutions for the Vlasov equation with non-uniform
number density, azimuthal rotation, possible accretion flows and
non-uniform temperature anisotropy have been constructed. For
doing this, the distribution function has been represented in
terms of first integrals and adiabatic invariants, as follows from
conservation laws of the single-particle dynamics, and taking into
account suitable kinetic constraints. As a consequence, based on
the perturbative kinetic approach, explicit constitutive equations
for the fluid fields have been determined, which are accurate to
first-order in the relevant expansion parameters. This permits the
construction of asymptotic solutions of the corresponding MHD
fluid equations systematically retaining all first-order
corrections, including effects due to FLR-diamagnetic and
energy-correction contributions.

Several physical issues have been analyzed. These concern: a)
deriving equilibrium equations of state for the species pressure
tensor components; b) establishing a fluid angular momentum
conservation law and comparing it with the predictions of standard
fluid treatments; c) investigating possible kinetic accretion
processes within equilibrium AD configurations (without perturbing
the equilibrium) and deriving a related kinetic accretion law; d)
demonstrating the existence of a quasi-stationary kinetic dynamo
mechanism for the self-generation of poloidal and toroidal
magnetic field.

Potential applications of the theory include, in particular, the
case of radiatively inefficient accretion flows. As mentioned
above, future work will address making stability analysis of the
kinetic equilibria presented here, so as to investigate further
mechanisms which could be responsible for driving the accretion
flow in the collisionless regime.

\textbf{Acknowledgments - }This work has been partly developed in
the framework of MIUR (Italian Ministry for Universities and
Research) PRIN Research Programmes and of the Consortium for
Magnetofluid Dynamics, Trieste, Italy.

\end{document}